\documentclass[pre,twocolumn,superscriptaddress,floatfix]{revtex4-1}  

\usepackage[paperwidth=210mm,paperheight=297mm,centering,hmargin=2cm,vmargin=2.5cm]{geometry}

\usepackage[pdftex]{graphicx}
\usepackage{amsmath,amsfonts,amssymb}
\usepackage{natbib}
\usepackage{MnSymbol}
\usepackage{csquotes}

\usepackage[]{algorithm}
\usepackage{algorithmic}

\usepackage{hyperref}
\usepackage{xcolor} 

\definecolor{bluemoi}{rgb}{0.25,0.50 ,0.75} 

\hypersetup{
  bookmarksopen=true,
  pdftitle=" ",
  pdfauthor=" ", 
  pdfsubject=" ", 
  pdftoolbar=true, 
  pdfmenubar=true, 
  pdfhighlight=/O, 
  colorlinks=true, 
  pdfpagemode=UseNone, 
  pdfpagelayout=SinglePage, 
  pdffitwindow=true, 
  linkcolor=bluemoi, 
  citecolor=bluemoi, 
  urlcolor=bluemoi 
}

\makeatletter
\renewcommand{\figurename}{\sf \textbf{Figure}}
\renewcommand{\thefigure}{\arabic{figure}}
\renewcommand{\fnum@figure}{\sf\textbf{\figurename}~\textbf{\thefigure}}
\renewcommand{\tablename}{\sf\textbf{Table}}
\renewcommand{\thetable}{\arabic{table}}
\renewcommand{\fnum@table}{\sf\textbf{\tablename}~\textbf{\thetable}}
\makeatother

\begin{document}

\title{Crowdsourcing the Robin Hood effect in cities} 

\author{Thomas Louail}
\thanks{Corresponding authors (thomas.louail@ifisc.uib-csic.es, maxime.lenormand@irstea.fr) who contributed equally to this work.}
\affiliation{Instituto de F\'isica Interdisciplinar y Sistemas Complejos IFISC (CSIC-UIB), Campus UIB, 07122 Palma de Mallorca, Spain}

\author{Maxime Lenormand}
\thanks{Corresponding authors (thomas.louail@ifisc.uib-csic.es, maxime.lenormand@irstea.fr) who contributed equally to this work.}
\affiliation{Irstea, UMR TETIS, 500 rue JF Breton, FR-34093 Montpellier, France}

\author{Juan Murillo Arias}
\affiliation{BBVA Data \& Analytics,Avenida de Burgos 16D, E-28036 Madrid, Spain}

\author{Jos\'e J. Ramasco}
\affiliation{Instituto de F\'isica Interdisciplinar y Sistemas Complejos IFISC (CSIC-UIB), Campus UIB, 07122 Palma de Mallorca, Spain}

\begin{abstract}
  Socioeconomic inequalities in cities are embedded in space and result in neighborhood
  effects, whose harmful consequences have proved very hard to counterbalance efficiently by
  planning policies alone. Considering redistribution of money flows as a first step toward
  improved spatial equity, we study a bottom-up approach that would rely on a slight evolution
  of shopping mobility practices. Building on a database of anonymized card transactions in
  Madrid and Barcelona, we quantify the mobility effort required to reach a reference situation
  where commercial income is evenly shared among neighborhoods. The redirection of shopping
  trips preserve key properties of human mobility, including travel distances. Surprisingly,
  for both cities only a small fraction ($\sim 5 \%$) of trips need to be altered to reach
  equality situations, improving even other sustainability indicators. The method could be
  implemented in mobile applications that would assist individuals in reshaping their shopping
  practices, to promote the spatial redistribution of opportunities in the city.
\end{abstract}

\maketitle


The growth of economic inequality has raised concern and attention in recent years
\cite{Piketty2014, Stiglitz2015}. In cities these inequalities are embedded in space, as a
result of entangled processes which include location choices of households and businesses,
daily mobility, segregation and closure attitudes, central planning, or global economic
restructuring. Over the course of several decades their joint actions have given rise to
segregated cities, characterized by uneven distributions of capital among their
neighborhoods. While the intensity of socioeconomic inequalities vary from one city to another,
the general observation that \enquote{some neighborhoods are poorer than others} has been made
for cities with different age, in every continent, and for different periods in urban history
\cite{Pinol2003, Goldsmith2010, Cassiers2012, UN-HABITAT2014}. An abundant literature has long
depicted the \emph{neighborhood effect} \cite{Friedrichs2003} -- the neighborhood impacts the
life trajectories of the residents, even when controlling for their individual characteristics
--, and highlighted its societal costs and enduring consequences \cite{Blau1982,
  Brooks-Gunn1997, Vallee2010, Womack1972, Chetty2015}.

\smallskip

Over the last decade, increasing volumes of digital geographic footprints have been produced by
individuals using mobile ICT devices, and these footprints have been increasingly analyzed by
scientists as well. These data are not free of biases \cite{Lewis2015} or privacy concerns
\cite{Montjoye2015}, but they undeniably constitute an important asset for understanding social
phenomena in detailed spatio-temporal contexts \cite{Lazer2009, Eagle2010, Onnela2007, Lu2012,
  Sun2013, Gonzalez2008}. They also have the potential to reveal the information required to
coordinate individuals' actions, so that large groups of people can tackle issues which are
distributed and spatial by nature. This is particularly true in the case of mobility networks,
which already integrate such footprints in feedback mechanisms: people produce data when
moving, and their travel decisions are partly guided by the data produced by others. Examples
include GPS navigation using real-time traffic data, local search and discovery of new places,
or location-based dating applications. So far, these footprints have been mainly used in
applications intended to enhance individual satisfaction (time savings, discovery of a
location, encounter of a partner), but they have also fostered spontaneous and large-scale
solidarity movements during disasters (e.g. Facebook's safety check, or the use of dedicated
Twitter hashtags). An important question is thus whether we can scale up, and address complex
issues through distributed and coordinated approaches relying on such data. Here we refer to
complex social issues for which improvements would necessarily occur on longer timescales.
There is a need to relate smart technology with sustainability and spatial justice in cities
\cite{McLaren2015}, and this implies building upon the existing practices of individuals. In
this work, we develop further this idea by focusing on a complex problem: the reduction of
spatial inequality in large cities.

\smallskip

\begin{figure*}[htbp]
  \centering
  \includegraphics[width=6.5in]{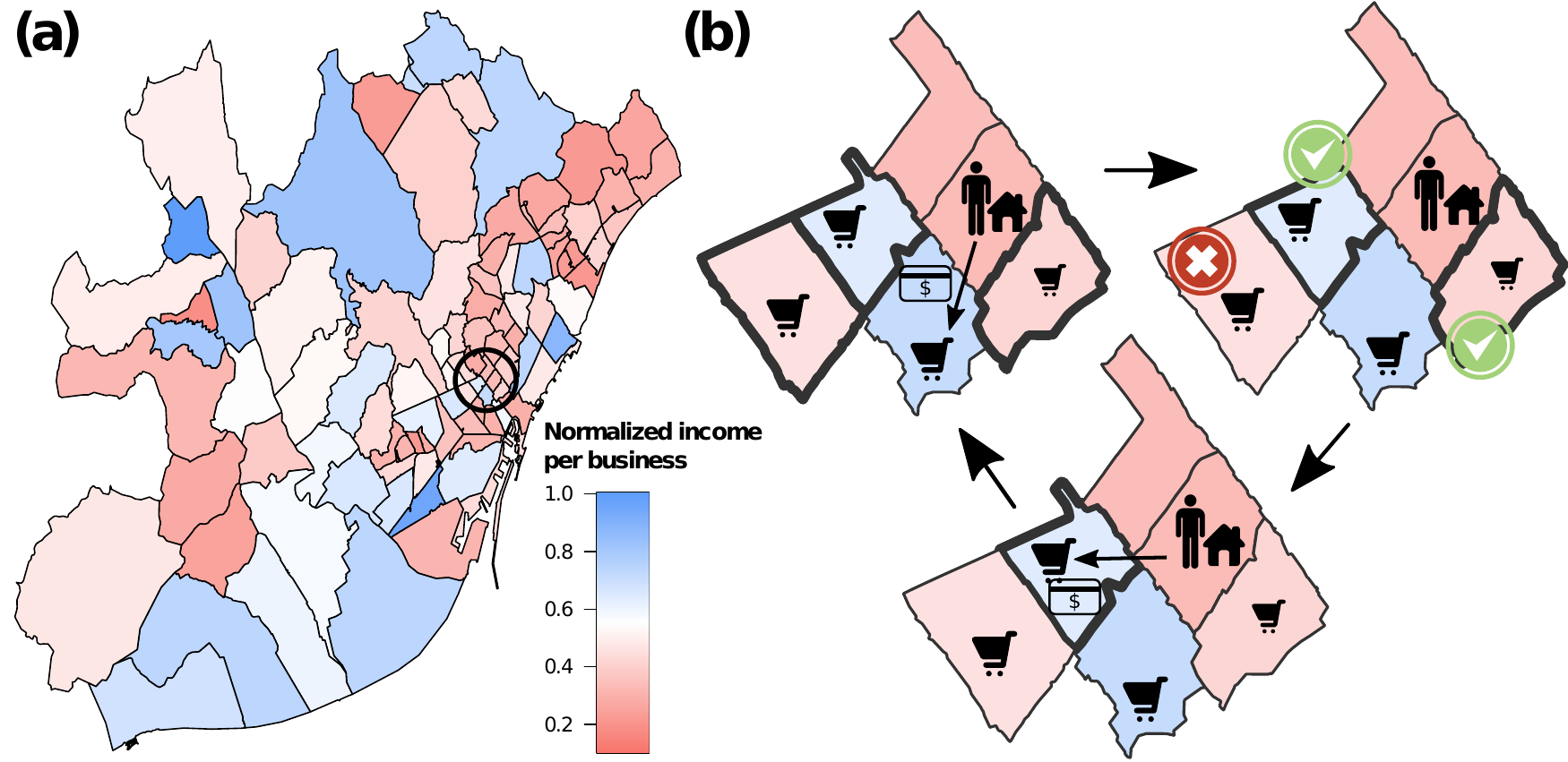}
  \caption{\label{fig:method}\textbf{Rewiring urban shopping trips.}
    (a) Average income per business in the neighborhoods of Barcelona
    resulting from individual transactions. The average income has
    been normalized by the maximum value among neighborhoods. The data
    correspond to $2011$ and is displayed by zip code in the
    metropolitan area. From this perspective, some neighborhoods are
    five times richer than others. (b) The general principle common to
    the iterative rewiring methods. At each step a transaction is
    randomly selected, along with the possible alternative businesses
    (highlighted in bold). If rewiring the transaction to one of them
    (randomly selected) decreases inequality between neighborhoods and
    matches the other constraints, then rewiring is performed.}
\end{figure*}

The \enquote{Robin-Hood effect} refers to a process through which capital is redistributed to
reduce inequality. A spatial and city-scale implementation would then consist in taking from
the rich neighborhoods to give to the poor. This role is normally played by the city's
governance, and is essential to mitigate spatial inequality. However, studies in cities
worldwide have demonstrated that top-down planning and fiscal policies alone seem inefficient
in significantly counterbalancing the numerous consequences of the neighborhood effect
\cite{Ostendorf2001, Satterthwaite2013}. It has also been long emphasized that developing
economic activity in disadvantaged regions indirectly benefits the surrounding populations, by
fostering job opportunities, transport facilities and increased safety \cite{Kanbur2005}. Here,
we study an original approach to rebalance economic activity among the neighborhoods of a
city. The scenarios we explore would not incur any additional environmental or monetary costs,
but would instead require slight modifications of daily shopping mobility practices.

According to surveys, shopping and leisure trips account for $15\%$ to $20\%$ of the
individuals' daily travels \cite{AASHTO}. Such trips virtually move money from one part of the
city to another, and directly contribute to shape the spatial distribution of wealth across
neighborhoods. By connecting areas, shopping trips also foster metropolitan integration and
``social cohesion'' \cite{Miciukiewicz2012}, whilst the resulting money flows are a key
component of the development of territories \cite{Ruault2014}. Large metropolitan areas are
characterized by mixed land use in many of their neighborhoods. Every time a resident has to
buy usual products such as food, gas or clothes, he/she can actually  choose among
several stores and neighborhoods to do so, sometimes without even increasing his/her travel
time.

In the following we focus on the two largest Spanish cities, Madrid and Barcelona. Performing
exploratory experiments, we demonstrate that they could be more evenly balanced thanks to the
cumulative addition of small and reasonable changes in a limited fraction of their residents'
shopping destinations. While there exists various spatial indicators for quantifying
territorial inequalities, static indicators fail to provide a clear picture of the collective
effort that would be required to reach a certain level of redistribution. With this in mind, we
quantify the proportion of individual shopping trips that should be redirected in order to
evenly share the commercial income between neighborhoods, the first step of a conceivable path
toward a spatial redistribution of opportunities in the city. We show that alternative mobility
scenarios not only allow to distribute money more evenly in space, but also to enhance the
spatial mixing of residents through their shopping mobility, without increasing the total
distance traveled, nor changing the individuals' effective purchases and mobility routines. The
following of this paper is an exploration on how money flows in cities could be more evenly
distributed, if shopping spatial behavior was slightly restructured. In the following of this
paper, we use zipcodes delimitation as proxies for neighborhoods. While zipcodes correspond
to administrative units which are very coarse-grained proxies for effective neighborhoods,
this choice was imposed by the spatial resolution of the available data.

\section*{Material and Methods}

\subsection*{Data}

We use a dataset containing the metadata of one year of bank card payments from more than
$150,000$ anonymous users in over $95,000$ businesses of Barcelona and Madrid. Each transaction
is time-stamped and contains the information collected by the bank on both the cardholder and
the business. It also includes the customer's age and residence's zipcode, the business
category and its geographical coordinates (see the Appendix for details and
\cite{Martinez:2016, Yoshimura:2016} for other recent examples of research relying on similar
data). From these data there are two obvious ways to estimate inequality among neighborhoods:
first, in measuring the income of their residents -- indirectly estimated through the amount of
money spent during the year; second, in measuring the income resulting from the commercial
activity of businesses located in these neighborhoods. The latter is particularly interesting
because it results from the spatial organization of shopping trips, which may be much easier to
alter than any other type of daily trips, notably commuting. The average commercial income of
businesses in Barcelona's neighborhoods, resulting from shopping trips, is mapped on Figure
\ref{fig:method}a. This map reveals that according to this measure, some neighborhoods are
indeed five times richer than others.

\subsection*{Rewiring the shopping trips networks}

From the data for both cities we construct the bipartite spatial network whose nodes are
individuals and businesses, and whose edges stand for transactions (see Figure S\ref{FigS2} in Appendix). We
then perform rewiring experiments, in which randomly selected transactions are redirected
toward alternative businesses of the same category, but located elsewhere in the city (Figure
\ref{fig:method}b). The rewiring methods we implemented operate directly at the level of
individual transactions (see Figure S\ref{FigS2} in Appendix). A rewiring operation then consists in randomly
selecting a transaction $t_{r,b}$ (made by user $r$ in business $b$), and an alternative
business $b'\neq b$, such than $b'$ and $b$ are of the same category (see in Appendix the details of
the 16 business categories), but located in different neighborhoods. The rewiring occurs only
if the change fulfills a number of constraints which are expressed at the city level. The
calculation of these constraints are based on the candidate configuration of the shopping trips
network (i.e. after $k+1$ rewiring operations), the current configuration (after $k$ rewiring
operations) and the original shopping trips network.

\subsubsection*{Four dimensions to assess the likelihood of the network's configuration}
\label{sec:metrics}

We consider four dimensions to assess the different network configurations from an economic,
social and environmental point of view. Since our main objective is to rebalance the
distribution of commercial income among the neighborhoods, we first focus on the economic
dimension. We denote by $W_k$ the \emph{wealth inequality among the city's neighborhoods} after
$k$ rewiring operations. It is defined as the distance to a reference homogeneous situation,
where the commercial income resulting from purchases would be equally shared among all
neighborhoods. We have:
	
\begin{equation}
  \label{eq:Wk}
  W_k = \sum_{i=1}^N (\overline{w}^i_k-w^*)^2 ,
\end{equation} 

where $N$ is the number of neighborhoods, $\overline{w}^i_k$ is the average income of the
businesses located in the neighborhood $i$ after $k$ rewiring operations, and $w^*$ represents
the wealth per neighborhood in the reference configuration where commercial income is evenly
distributed across neighborhoods, such that

\begin{equation}
  \label{eq:w}
  w^*=\frac{1}{N}\sum_{i=1}^N \overline{w}^i_k.
\end{equation}	

Another important aspect is related to the social nature of mobility in the city that might
prevent some neighborhoods from ghettoization. To measure to what extent individuals residing
in various neighborhoods mix in the city space as a result of their travels, for each
neighborhood $i$ we count the number of times $(s^{i1}_k,...,s^{iN}_k)$ the residents of $i$
traveled to each of the $N$ neighborhoods ($i$ included), after $k$ rewiring operations. Then,
by averaging the vector of trips over all the neighborhoods, we compute a geographical
diversity index $S_k$ (after $k$ rewiring operations),

\begin{equation}
  S_k = \frac{1}{N}\sum_{i=1}^N \sum_{j=1}^N (s^{ij}_k-s_i^*)^2 , \label{S}
\end{equation}

where $s_i^*$ represents the homogeneous distribution of visits originating from $i$ and in
direction to all neighborhoods,

\begin{equation}
  \label{eq:w2}
  s_i^*=\frac{1}{N} \sum_{j=1}^N s^{ij}_k.
\end{equation}

The third considered dimension is the distance traveled by individuals. Summing the distances
traveled by individuals for all their shopping trips, we can compute $D_k$ the total distance
traveled, as measured after rewiring $k$ transactions. Details about the method used to
estimate the shopping trips distances are available in Appendix.

Finally, we are also interested in individual mobility routines and the tendency of individuals
to return to already visited places. For each individual we calculate an exploration rate
$\rho_k$. It is defined as the number of unique businesses he/she has visited divided by
his/her total number of transactions, after $k$ rewiring operations. Considering the empirical
peaked distribution of $\rho_k$ among the population of customers (see Figure
\ref{fig:mobility-indicators}b), in the following we only consider the average value
$\bar{\rho}_k$.

\subsubsection*{Rewiring constraints}
\label{sec:constraints}

As mentioned above, a candidate reconfiguration of the shopping trips network
  ($k+1$ rewiring operations) will occur if and only if the proposed change respects a number
  of constraints regarding the current configuration ($k$ rewiring operations) and the original
  shopping trips network. We consider four constraints, each of them concerns one of the four
  economic, social and environmental dimensions described in the previous section,

\begin{itemize}
\item A first constraint applies on the \emph{wealth distribution}; it ensures that each
  destination change contributes to iteratively homogenize the distribution of commercial
  income across neighborhoods.
\item A constraint on the \emph{spatial mixing} of individuals resulting from their
  shopping travels. In order to be accepted, a rewiring operation has to preserve the diversity
  of neighborhoods visited, hence the degree of spatial mixing of individuals residing in
  different neighborhoods.
\item A third constraint on the \emph{total distance traveled}, to guarantee that each
  destination change does not result in increasing the total distance traveled. The distance
  associated to each individual transaction is calculated with regard to the individual's main
  activity place at this moment of the day.
\item Finally, a constraint on the spatial \emph{exploration rate} of individuals, to
  preserve the behavioral mobility routines measured in the population.
\end{itemize}

All constrains have the same form and are satisfied if the following condition
holds

\begin{equation}
  \label{eq:general-form}
  X_{k+1} \leq
  \begin{cases}
    X_{k} & \text{if } X_{k+1} \geq \alpha X_0,\\
    \alpha X_0 & \text{otherwise},
  \end{cases}
\end{equation}

\noindent where $k$ denotes the number of rewiring operation, and $\alpha$ is a parameter
positive or equal to zero. The general form of Equation \ref{eq:general-form} allows us to fix
an objective upper bound for each dimension $X_k$ with respect to its original value
$X_0$. Then as long as $X_k$ is greater than $\alpha X_0$, each rewiring operation must
decrease $X$. Once $X_k$ is smaller than $\alpha X_0$, then rule \ref{eq:general-form} ensures
that none of the following rewiring operations will increase $X_k$ above $\alpha X_0$. An
experiment is then defined by a set of four values
$(\alpha_W, \alpha_S, \alpha_D, \alpha_{\bar{\rho}})$ that specify the maximal value desired
for each variable of interest.

\begin{figure*}[htbp]
  \centering
\includegraphics[width=6.5in]{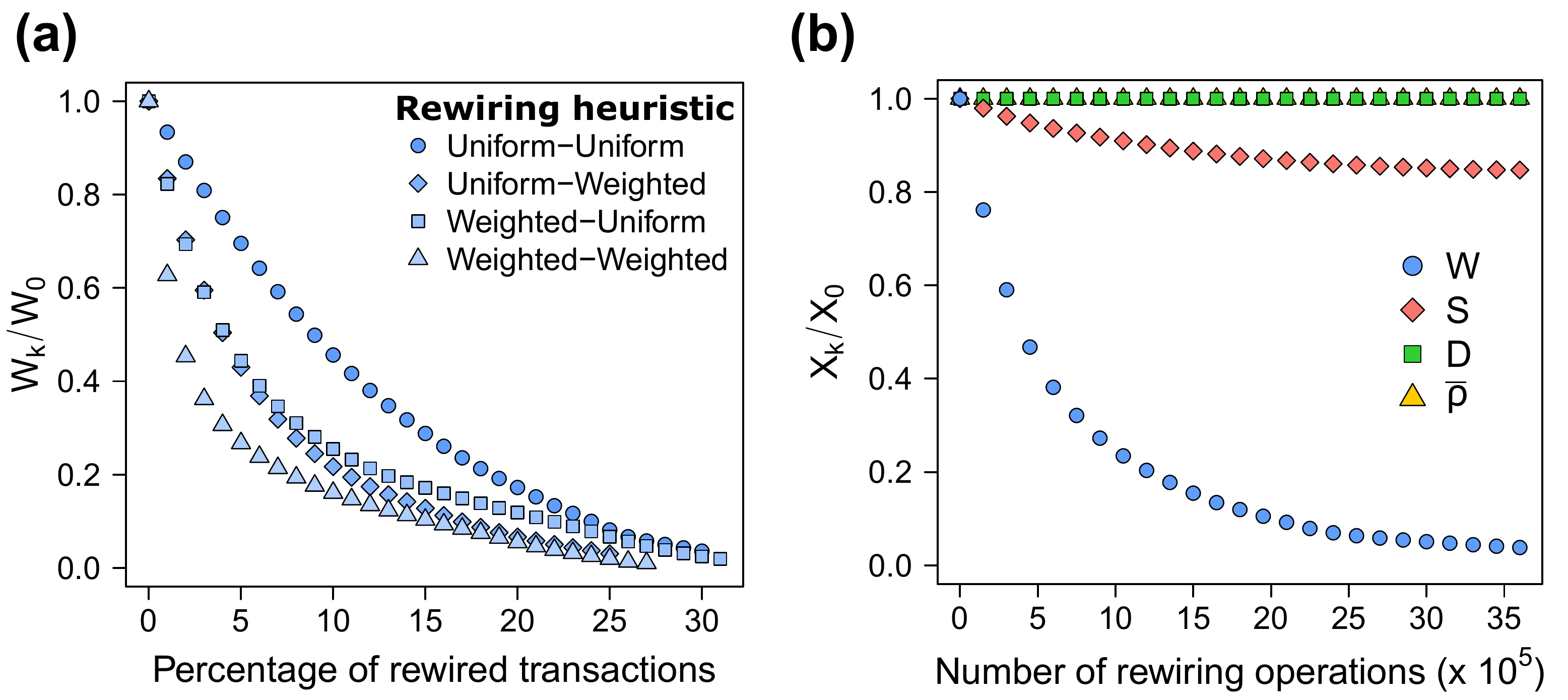}
\caption{\label{fig:reachability} \textbf{Decreasing spatial inequality in the city by adapting
    daily shopping destinations.}  (a) Decrease of wealth inequality among neighborhoods as a
  function of the fraction of transactions rewired, for various rewiring methods. Four
  combinations of choice heuristics are considered, \enquote{Uniform-Uniform},
  \enquote{Uniform-Weighted}, \enquote{Weighted-Uniform} and \enquote{Weighted-Weighted}. (b)
  Decrease of wealth inequality ($W_k/W_0$) while preserving the spatial mixing index
  ($S_k/S_0$), the total distance traveled ($D_k/D_0$) and the exploration rate
  ($\bar{\rho}_k/\bar{\rho}_0$), as a function of the number of rewiring operations. Values
  have been averaged over hundreds of replications. The bars represent the minimum and the
  maximum values obtained but in most cases are too close to the average to be seen (see Figure
  S\ref{FigS10}-S\ref{FigS11} in Appendix for Madrid).}
\end{figure*}

\subsubsection*{Algorithm}
\label{sec:algo} 

The process is unambiguously specified by Algorithm \ref{algo}. One should note that the
products purchased and the amount of expenses of each individual are preserved. This iterative
process is run until the rewiring rate falls below a given threshold (see the Appendix for more
details). Since the rewiring process is stochastic, all the results we discuss in the following
sections have been averaged over hundreds of replications. Numerous rewiring methods fulfilling
the aforementioned conditions could be proposed. However, we favored a numerical approach
because of the large number of transactions ($\sim 10^7$) and also because of the constraints
we impose to guarantee realistic and interesting properties.

Besides, the random selection of (i) the transaction to rewire and (ii) of the candidate
business (step 1. of the algorithm) can be uniform (denoted \enquote{Uniform} sampling
hereafter) or proportional to their amount in the case of transactions and/or inversely
proportional to the average income of the targeted neighborhood for the businesses (denoted
\enquote{Weighted} sampling hereafter). Even more informed methods might be proposed, but for
the sake of simplicity only simple random procedures are tested in the following.

\begin{algorithm}[H]
  \caption{Rewiring the shopping trips network}
  \label{algo}
  \begin{algorithmic}
    \STATE \textbf{Input:} Shopping trips network obtained after $k$ operations
    \STATE \textbf{1.} Pick at random a candidate $(k+1)^{th}$ rewiring operation:
    \STATE \hspace{0.75cm} Pick at random a transaction $t_{r,b}$ 
    \STATE \hspace{0.75cm} Pick at random a business $b'$ 
    \STATE \textbf{2.} Compute ($W_{k+1}$,$S_{k+1}$,$D_{k+1}$,$\bar{\rho}_{k+1}$)
    \STATE \textbf{3.} \textbf{if} for each dimension $X\in\{W,S,D,\bar{\rho}\}$ \textbf{we have}
    \STATE $$
  				X_{k+1} \leq
  				\begin{cases}
    				X_{k} & \text{if } X_{k+1} \geq \alpha X_0,\\
    				\alpha X_0 & \text{otherwise},
  				\end{cases}
			$$
    \STATE \textcolor{white}{\textbf{3.}} \textbf{then} Accept the rewiring operation
    \STATE \textcolor{white}{\textbf{3.}} \textbf{else} Return to step \textbf{1.} 
  \end{algorithmic}
\end{algorithm}

\section*{Results}
 
\subsection*{Reachability of even spatial distributions} 

We first investigate the reachability of an even spatial distribution of the commercial income
resulting from individual purchases, while the variables $S$, $D$ and $\bar{\rho}$ remain in
the range of their empirical values. To address this question, we apply the rewiring method
previously described with the four constraints of Equation \ref{eq:general-form} such as
$\alpha_W=0$, $\alpha_S=1$, $\alpha_D=1$ and $\alpha_\rho=1$. This constitutes our
\textit{Reference} scenario. Figure \ref{fig:reachability}a shows the evolution of inequality
in the urban area of Barcelona as a function of the fraction of rewired transactions, according
to various sampling methods. Surprisingly, even with basic random sampling methods, it is
possible to reduce spatial inequality between neighborhoods by more than $80\%$ while
reassigning only $20\%$ of individual transactions. All the methods produce the same
qualitative behavior -- an early regime of very fast decay, followed by a regime of slower
decay. Weighted methods are naturally more efficient, and allow to reach spatial equity by
redirecting a smaller fraction of transactions. In particular, a reduction of wealth inequality
of $80\%$ ($W_k/W_0$) can be obtained by rewiring only $5\%$ of the transactions if the
sampling method is double weighted.

The state of the other variables $S$, $D$ and $\rho$ is also monitored along the process, as
shown in Figure \ref{fig:reachability}b for a Uniform-Uniform sampling method. What makes the
previous results remarkable is in fact that income redistribution is achieved without
increasing the distance traveled by individuals ($D$), nor changing their mobility routines
($\bar{\rho}$). Moreover, a positive side-effect is to increase the frequency of encounters of
individuals living in different parts of the city -- as indicated by the decrease of $S_k/S_0$
--, an effect that could not be anticipated from the rewiring constraints alone
($\alpha_S=1$). The increase of spatial mixing is the consequence of individual shopping trips
more evenly distributed in the city space, required to homogenize the income among
neighborhoods. The behavior of $S$ is non trivial, notably because one could imagine
unrealistic solutions that would simultaneously even the spatial distribution of business
income and decrease the total distance traveled, by rewiring most of the shopping trips to the
closest neighborhood containing businesses of the relevant category. In this case the spatial
mixing of individuals would decrease dramatically, and $S$ the distance to an homogeneous
mixing would increase. Here the decay of $S/S_0$ guarantees that it is not the case.

\subsection*{Preservation of human mobility properties} 

We wish to control further the likelihood of the rewired shopping mobility networks, and ensure
that they preserve the spatial properties of individual human mobility. A small set of
indicators have proved to be useful to describe the statistical and spatial properties of human
mobility \cite{Gonzalez2008}. These include the jump length between consecutive locations
$\Delta_r$, the radius of gyration $r_g$ (with
$r_g(i) = \sqrt{\frac{1}{N}\sum_{j=1}^{N}(x_j(i)-x^*(i))^2}$, $r_g(i)$ is the characteristic
travel distance of individual $i$ with center of mass $x^*(i)$ after $N$ displacements), and
the tendency to return to already visited places $\rho$. In our case, for each individual
$\rho$ is simply defined as the ratio between the number of unique businesses visited and the
total number of transactions.

\begin{figure}[!h]
  \centering
  \includegraphics[width=8.25cm]{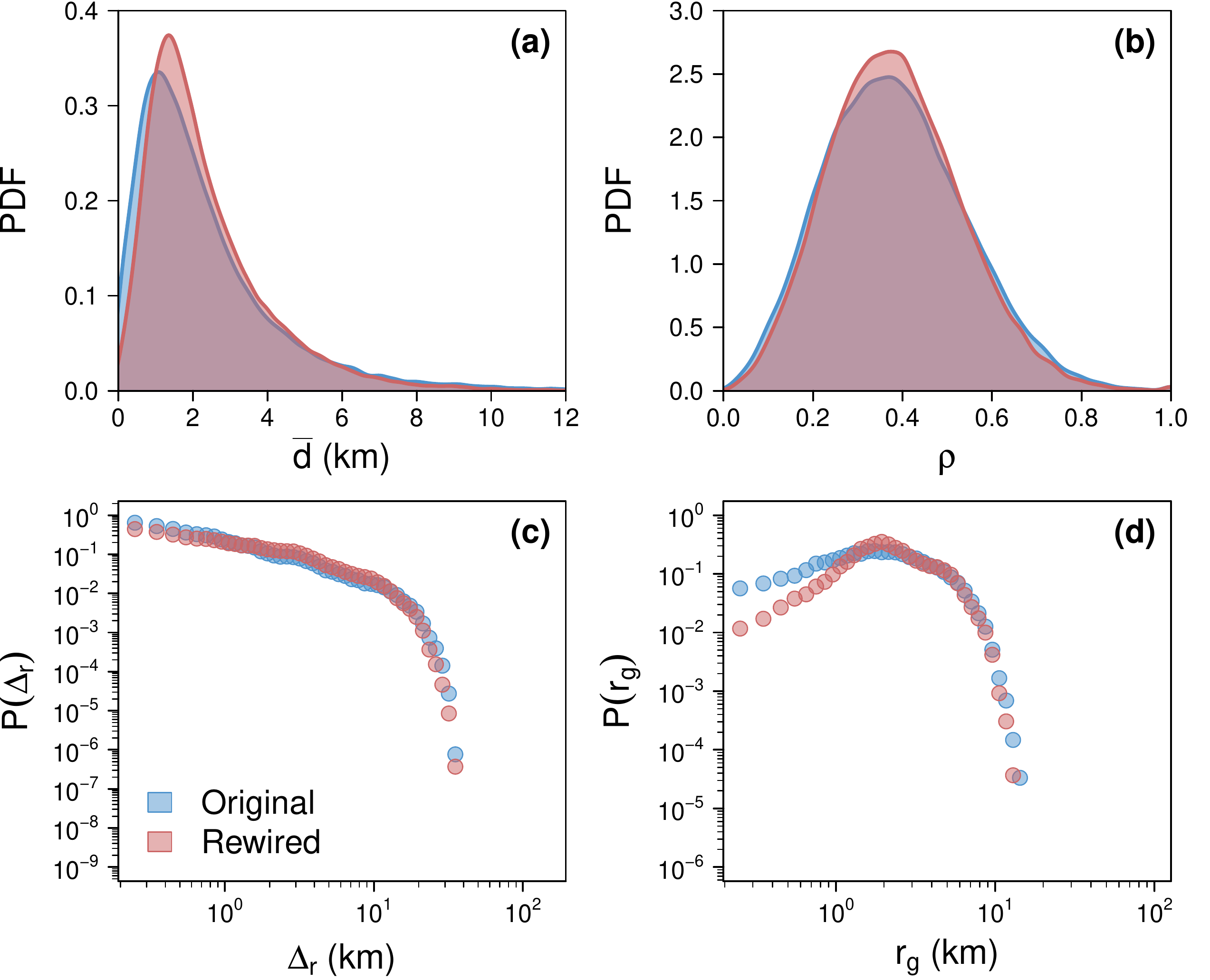}
  \caption{\label{fig:mobility-indicators} \textbf{Observed and simulated distributions of
      human mobility indicators.}  The distribution of jump lengths $\Delta_r$, the radius of
    gyration $r_g$, the tendency to return to already visited places ($\rho$) and the
    individual average distance traveled ($\bar{d}$) are considered. Values measured on the
    empirical data are in blue, while those obtained after rewiring are in red. The calculation
    of $\Delta_r$ and $r_g$ is based on the business' exact geographical coordinates. The
    simulated distributions plotted here correspond to one particular replication, see Figure
    S4 for the robustness of the results and Figure S\ref{FigS12} in Appendix for the same curves for Madrid.}
\end{figure}

Figure \ref{fig:mobility-indicators} shows their empirical and simulated values, plus the
average distance $\bar{d}$ traveled by each individual for each shopping trip (see the Appendix
for details on the calculation of shopping trips distances). On each panel both curves
overlap almost perfectly, indicating that the rewiring has no significant effect on the key
mobility properties. The simulated distributions of $\bar{d}$ and $r_g$ are slightly more
peaked than the empirical ones.

We showed in a previous study that young adults tend to spend their money further from their
neighborhood of residence \cite{Lenormand2015}. Coherently their shopping trips are those that
are the most affected in the simulated scenarios (see Figure S\ref{FigS6} in Appendix). But the simulated
scenarios also contain some questionable aspects. For example, elderly people are those whose
displacements would increase the most (with regard to their current shopping travel distances --
see Figure S\ref{FigS5} in the Appendix).

\subsection*{Multi-objective improvement}

\begin{figure}
  \centering
  \includegraphics[width=8.25cm]{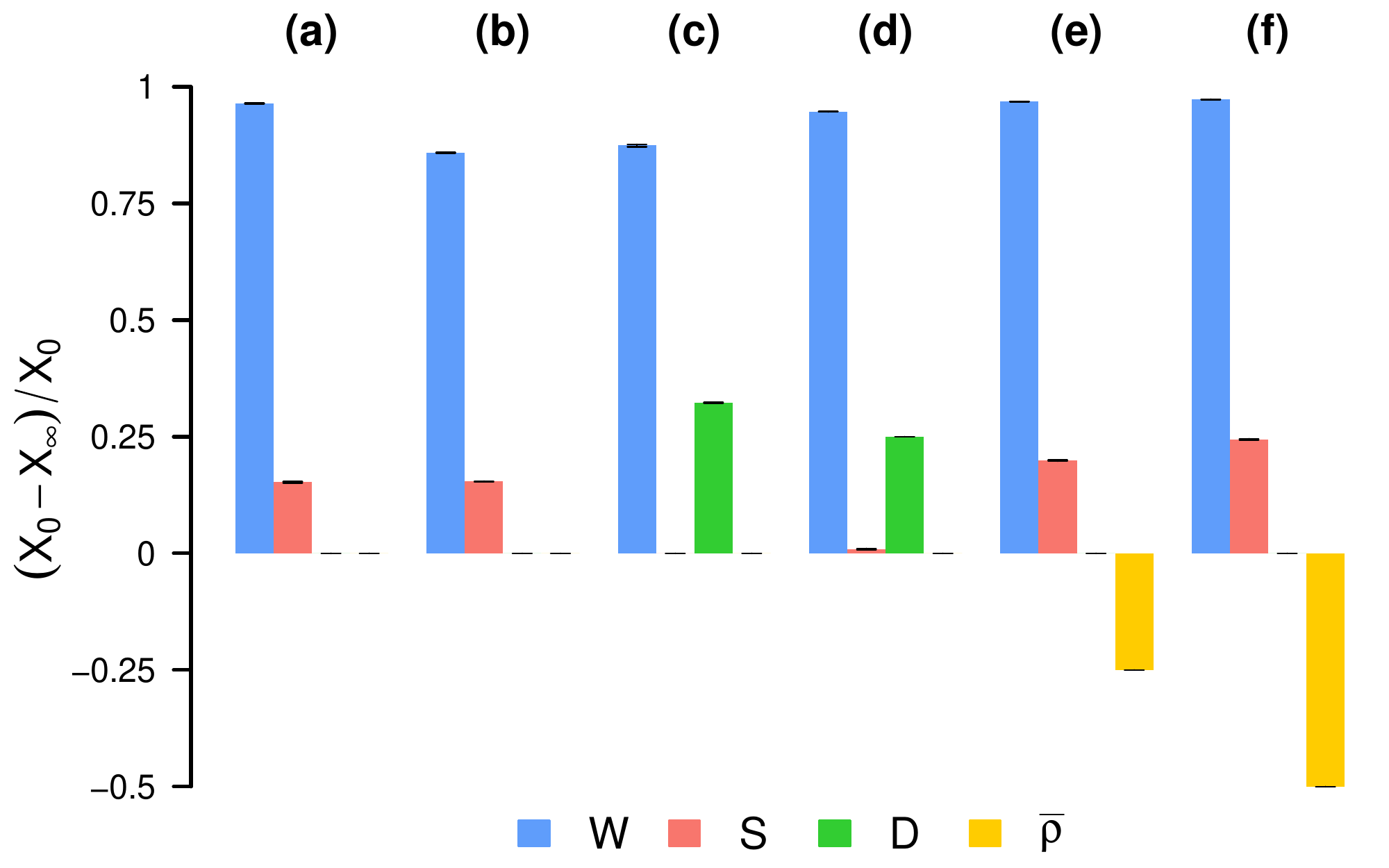}
  \caption{\label{fig:exploration}\textbf{Multi-criteria improvement of shopping mobility.}
    Each group of bars gives the relative gains or losses for the four indicators $W$, $S$, $D$
    and $\bar{\rho}$. Experiments are described in Table \ref{tab:experiments}. See Figure S\ref{FigS13}
    in Appendix for Madrid.}
\end{figure}

We now perform multi-criteria rewiring experiments in order to measure to what extent
redistribution can be achieved while improving simultaneously other important aspects of urban
mobility. To this end, we perform the series of experiments summarized in Table
\ref{tab:experiments}. The objective is to even the wealth distribution among neighborhoods
($\alpha_W=0$) and also improve either $S$, $D$ or $\bar{\rho}$ without worsening the other
two. Figure \ref{fig:exploration} gives the relative gains and losses upon the four indicators,
and the last two columns of Table \ref{tab:experiments} contain the asymptotic values obtained
for the reduction rate of wealth inequality, for Barcelona (B) and Madrid (M).

\begin{table}[!h]
  \caption{\label{tab:experiments} \textbf{Experiments performed.} 
    Column W indicates the relative gain of $(W_0 - W)/W_0$. 
    The first value is for Barcelona (B) and the second for Madrid (M). }
  \begin{center}
    \begin{tabular}{llllll}
      \hline 
      \textbf{Experiment} & $\alpha_W$ & $\alpha_S$ & $\alpha_D$ & $\alpha_{\bar{\rho}}$ & W (B/M)\\
      \hline 
      (a) Reference & $0$ & $1$ & $1$ & $1$ & $96.4\%/99.5\%$\\
      (b) Spatial mixing $\uparrow$ & $0$ & $0.75$ & $1$ & $1$ & $85.9\%/78.1\%$ \\
      (c) $50\%$ energy savings & $0$ & $1$ & $0.5$ & $1$ & $87.4\%/84.8\%$\\
      (d) $25\%$ energy savings & $0$ & $1$ & $0.75$ & $1$ & $94.7\%/98.8\%$\\
      (e) Exploration rate $\uparrow$ & $0$ & $1$ & $1$ & $1.25$ & $96.8\%/99.9\%$\\
      (f) Exploration rate $\uparrow\uparrow$  & $0$ & $1$ & $1$ & $1.5$ & $97.3\%/100\%$ \\
      \hline
    \end{tabular}
  \end{center}
\end{table}

These experiments proove that it is not always possible to combine significant improvements on
several dimensions simultaneously. This is not an issue with the method, but rather with the
set of objectives which are somewhat opposite. Most individuals perform their shopping trips
near their residence -- as highlighted by the empirical distributions in Figure
\ref{fig:mobility-indicators} -- and consequently it is not feasible to diversify the
neighborhoods where an individual regularly travels to -- in order to improve spatial mixing
$S$ -- and at the same time decrease the total travel distance $D$. More surprisingly,
experiment (b) indicates that it is also not possible to simultaneously improve the wealth
redistribution and the spatial mixing of individuals. The two indicators are based on different
metrics (the amount of money spent per business for $W$ and the number of trips for $S$), which
imply different reference egalitarian situations (see Figure S\ref{FigS7} in Appendix for more
details). Optimization is thus a trade-off between the various consequences of shopping
mobility at the city scale. However, experiments (c) and (d) prove that it is possible to
significantly decrease the total distance traveled and in the same time to strongly reduce
wealth inequality among neighborhoods, but not as much as in the reference experiment.  Still,
it is remarkable that experiment (d) results in an alternative mobility network such that the
spatial inequality of the average business income is reduced by $95\%$, while the total
distance associated to shopping mobility is reduced by $25\%$, the level of spatial mixing is
preserved, as well as the individual mobility routines. Finally, experiments (e) and (f) show
that even if residents deeply restructured their mobility routines, and typically started going
to a new business each time they perform a new shopping trip, keeping control of the total
distance traveled in the city would prevent from increasing the mixing of individuals coming
from different neighborhoods beyond $25\%$. The gains in terms of wealth redistribution would
not be significant when compared to the reference experiment (a).

\section*{Discussion}

Reducing urban segregation and increasing spatial justice are some of the major challenges
faced by cities worldwide, and the digital footprints passively produced by their residents
constitute a promising resource to help addressing these issues from the bottom. This study is
a first attempt to quantify the relation between shopping mobility and the spatial distribution
of economic activity in the city. The alternative shopping trips resulting from our experiments
offer an interesting trade-off between the preservation of essential aspects -- the effective
purchases of individuals and households, and their mobility properties -- and some reasonable
changes in the places where they spend their money. The addition of small changes in the
shopping destinations of individuals can dramatically impact the spatial distribution of money
flows in the city, and the frequency of encounters between residents of different
neighborhoods, even if the total number of changes remains relatively small. These results have
important consequences, and they lead in particular to the decisive question of the effective
implementation of alternative shopping travels, like those drawn by our experiments. While the
decision process behind each individual redirection may appear intricate for a single person,
one could easily imagine dedicated mobile applications, querying databases similar to the one
we used in this paper. Their purpose would be to assist their users in a transition toward a
more socially and spatially concerned shopping mobility.

\subsection*{Limitations of the study}

However, one should keep in mind that individuals do not guide most of their travel decisions
by philanthropy, but instead by balancing accessibility, price and business
characteristics. Individuals first choose their casual shopping destinations with regard of
transport facilities and travel time budget \cite{Recker1978}. Here as accessibility
information we considered the Euclidean distance between neighborhoods. However, in urban
environments the Euclidean distance is rarely a direct proxy for travel time
\cite{Gallotti2015}, and some of the rewired shopping trips are unlikely to be performed in the
real world. more accurate measures of travel time. In future studies transport APIs and road
network data could be used instead to calculate more realistic travel distances between points
in the city. Also, from a choice point of view, distance is not the only one determinant of
shopping destination choices. The inequality might be also due to the inequality between
businesses of the same category, which may differ substantially in terms of the price, product
quality, etc. The reader should also remind that prices of retail goods are not uniform across
a city's neighborhoods, they may display strong spatial variation, affecting consumer choices
and spatial behavior in complex ways.

We also assumed that every shopping trip follows the simple pattern
$A\rightarrow B\rightarrow A$, and we did not consider the more complicated case of chained
trips (e.g.  $A\rightarrow B\rightarrow C\rightarrow A$) during which individuals join several
trips associated with different purposes \cite{Schneider2013}. Destination choices for certain
types of shopping travels are also motivated by reasons that are not only related to the sole
proximity (e.g. the local market is a place where residents can build a sense of community,
discuss the problems of the neighborhood, make social contacts useful in everyday
life). People also tend to choose the places where they spend money according to several other
key factors, the price of products in the first place, but also according to some more personal
appreciations, such as the "atmosphere" of neighborhoods and the feeling of well-being they
provide. In large cities, the neighborhoods strongly differ in the quality of their planning
and architecture, in their public spaces, in their amenities and leisure opportunities,
commercial fabric, in their safety. Additionally, the changes might be considered as
problematic, since a profound spatial reorganization of shopping mobility in the city could
have consequences on the spatial structure of employment in the first place (see Figures S\ref{FigS8} and
S\ref{FigS9} in Appendix showing the evolution of the number of clients and transactions per business), then on
residences, finally on their commercial offer and ambiance. This questions the likelihood and
desirability of the objective configuration we chose, in which the average commercial income
per business is evenly balanced across neighborhoods. Still it has the advantage to be
unambiguously defined, and constitutes an immediate, easy-to-think-with reference situation

For the sake of simplicity, we considered none of the aforementioned factors but they could be
implemented in more involved frameworks derived from our work. A number of additional
constraints, aiming at making the rewiring schemes more realistic from a spatial economy
perspective, could be introduced in future scenarios. These include the preservation of the
number of transactions and income per business (or at least a certain fraction of them); the
restriction of the rewiring operations to certain business categories; limiting changes for
certain socio-demographic categories of population (such as elderly people); or ensuring the
temporal likelihood of the simulated scenarios (so that the rerouted shopping travels are
homogeneously distributed through time).

\subsection*{Concluding remarks}

There are recent, encouraging examples of fast and wide adoptions of new daily practices, whose
benefits are essentially collective. Examples include garbage differentiation
\cite{Robinson2005}, the increasing role of bicycle in urban transport and the development of
bicycle sharing systems \cite{Dill2003}, or the open-source movement and the dedication of a
growing number of individuals to collectively build free knowledge databases (e.g. Wikipedia,
StackExchange, the free software movement). In these cases, the remuneration of participants,
if any, is essentially symbolic. The success of a spatial counterpart to these altruistic
behaviors could rejuvenate the very meaning of the so-called sharing economy
\cite{McLaren2015}. As citizens produce the data that document their location and activity
patterns, in return these data could serve not only the specific interest of the institution
collecting them, but also support fair socio-economic initiatives. Our study brings evidence
that these geographical footprints we passively produce can support bottom-up responses to big
societal issues, an expected feature of truly smart cities.

\vspace*{0.5cm}
\section*{Acknowledgements} 

We wish to thank Julie Vall\'ee and Suma Desu for providing feedback
on an early version of the manuscript. We also thank Jean-Loup
Guillaume for fruitful discussions on the rewiring problem. Partial
financial support has been received from the Spanish Ministry of
Economy (MINECO) and FEDER (EU) under the project INTENSE@COSYP
(FIS2012-30634), and from the EU Commission through projects EUNOIA
and INSIGHT. The work of TL has been funded under the PD/004/2015, from the
Conselleria de Educaci\'on, Cultura y Universidades of the Government
of the Balearic Islands and from the European Social Fund through the
Balearic Islands ESF operational program for 2013-2017. JJR
acknowledges funding from the Ram\'on y Cajal program of MINECO.

\bibliographystyle{unsrt}
\bibliography{Crowdsourcing}

\onecolumngrid
\vspace*{2cm}
\newpage
\twocolumngrid

\makeatletter
\renewcommand{\fnum@figure}{\sf\textbf{\figurename~\textbf{S}\textbf{\thefigure}}}
\renewcommand{\fnum@table}{\sf\textbf{\tablename~\textbf{S}\textbf{\thetable}}}
\makeatother

\setcounter{figure}{0}
\setcounter{table}{0}
\setcounter{equation}{0}

\section*{Appendix}

\subsection*{Data preprocessing}

\begin{figure}[!h]
  \centering
  \includegraphics[width=\linewidth]{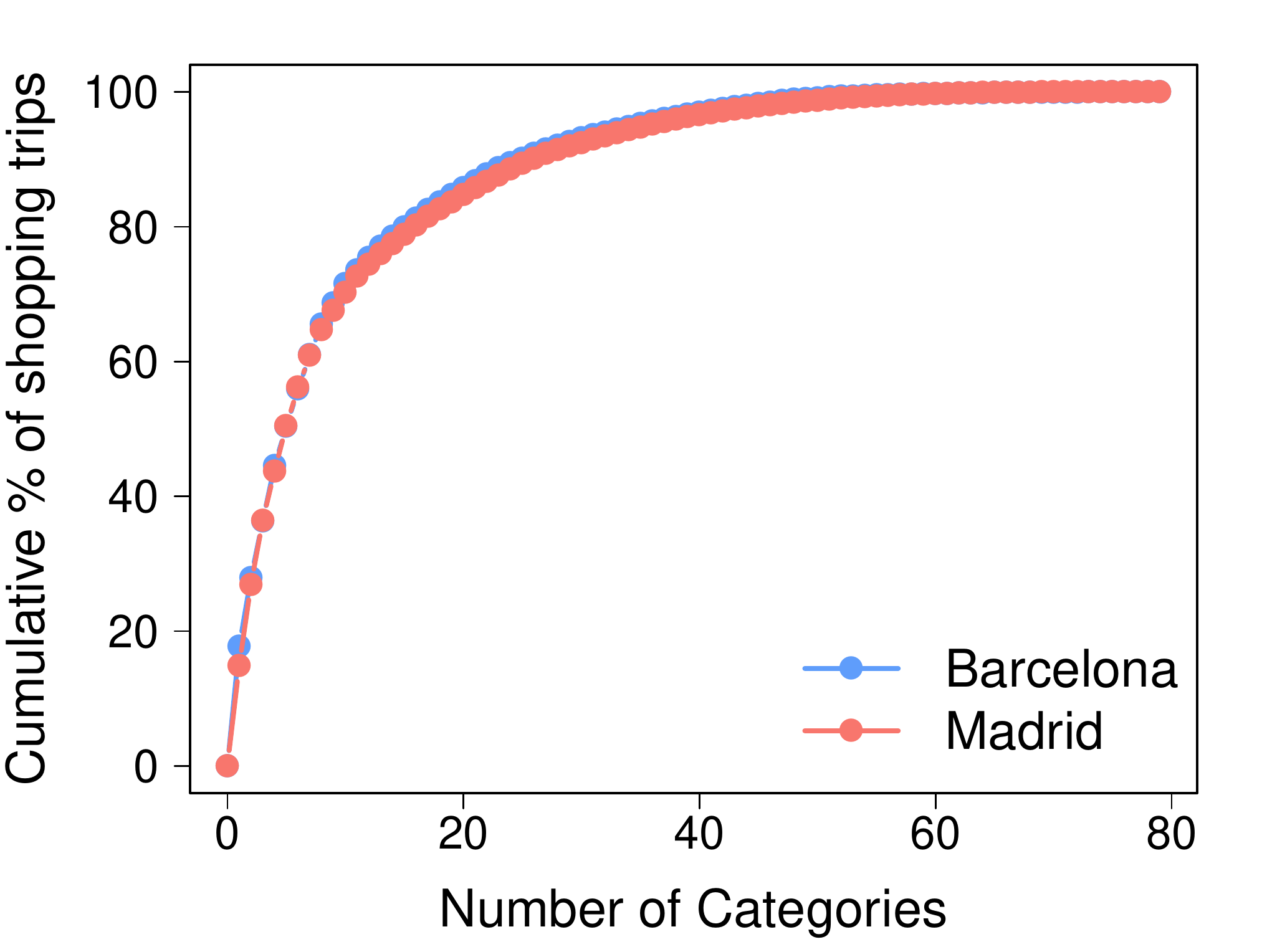}
  \caption{\label{fig:subcat} \textbf{Cumulative proportion of shopping
      trips as a function of the number of categories.} In blue the
    metropolitan area of Barcelona; In red the one of Madrid}
\end{figure}

The dataset contains information about $14$ million bank card
transactions made by customers of the Banco Bilbao Vizcaya Argentaria
(BBVA) in the metropolitan areas of Barcelona and Madrid in
$2011$. For both case studies, we only consider the credit card
payments whose amount was inferior to $1000$ euros, and which were
made inside the metropolitan areas, by bank customers that lived and
worked in the metropolitan area in $2011$. Each transaction is
characterized by its amount (in euro currency) and the time when the
transaction has occurred. Each transaction is also linked to a
customer and a business. Customers are identified with an anonymized
customer ID, connected with sociodemographic characteristics (gender,
age and occupation) and their postcode of residence. In the same way,
businesses are identified through an anonymized business ID, a
business category id, and the geographical coordinates of the credit
card terminal. Since we are primarily interested in daily shopping
mobility, we chose to consider the business categories that account
for the top 90\% of the daily shopping trips (see
Figure S\ref{fig:subcat}). The proportions of shopping trips associated
to each of the $20$ business categories we selected are available in
Table S\ref{tab:subcat}.

\begin{table}[!h]
  \caption{Proportion of shopping trips associated to each of the 20 business categories selected.}
  \label{tab:subcat}
  \begin{center}
    \begin{tabular}{lcc}
      \hline
      \centering Category   & Barcelona &  Madrid             \\
      \hline
      Supermarket	&	17.71	&	14.84	\\
      Hypermarket	&	10.25	&	12.09	\\
      Gas Stations	&	8.41	&	9.49	\\
      Restaurants	&	8.20	&	6.73	\\
      Retail store	&	5.84	&	2.82	\\
      Clothing store chain	&	5.48	&	4.67	\\
      Clothing store 	&	5.16	&	7.33	\\
      Pharmacy, optical and orthopedics	&	4.52	&	3.81	\\
      Department store	&	3.14	&	5.73	\\
      Hair and beauty	&	2.88	&	2.72	\\
      Electronics, computers and appliances	&	1.97	&	1.44	\\
      Bars and caf{\'e}	&	1.85	&	1.60	\\
      Shoe store	&	1.71	&	1.43	\\
      Toys and sports articles	&	1.43	&	1.33	\\
      Bookshop, music shop and stationery	&	1.42	&	1.04	\\
      Fast food restaurants and chains	&	1.13	&	2.38	\\
      Car dealership and garage	&	1.02	&	1.01	\\
      Bazaar	&	1.01	&	1.06	\\
      DIY store	&	0.99	&	1.08	\\
      Hospitals, clinics and doctors	&	0.91	&	0.88	\\
      \hline
    \end{tabular}
  \end{center}
\end{table}

\subsection*{Formal description of the rewiring process}
\label{sec:method}

From the data we extract $G(R, B, T)$ the bipartite network of all
credit card transactions performed by the city residents in businesses
located in the city, during the entire year. $R$ is the set of
residents, $B$ the set of businesses and $T$ the set of
transactions. Table S\ref{tab1} contains the characteristic attributes
of the network in the two cities studied. Each city is partitioned in
$N$ spatial units/neighborhoods (here the units correspond to zip
codes) and the network $G$ is spatial: each resident and each business
is located in one neighborhood. We denote $R_i$ (resp. $B_i$) the set
of residents (resp. businesses) located in the neighborhood $i$. The
sets are disjointed and we have $R = \cupdot R_i$ and $B=\cupdot B_i$,
with $i\in 1..N$. Additionally the businesses are also partitioned in
$C$ categories according to the products they sell, and we have
$B= \cupdot B_c, c \in 1..C$. The edges of the network represent the
card transactions, hence implicitly the shopping trips. We note
$t_{r, b}^{k}$ the k-th transaction performed by resident $r$ in
business $b$, and by $w(t_{r, b}^{k})$ its amount.

\begin{table}[!h]
  \caption{\textbf{Summary statistics of the two metropolitan areas and of the two transactions networks.}}
  \label{tab1}
  \begin{center}
    \begin{tabular}{lcc}
      \hline
      \centering Statistics         & Barcelona  &  Madrid      \\
      \hline
      Number of neighborhoods           & 97         &  123         \\
      Number of inhabitants (2009)  & 3,218,071  &  5,512,495   \\
      Area (km$^2$)                 & 634        &  1,935       \\
      Number of customers           & 42,023     &  118,447     \\
      Number of businesses          & 40,618     &  55,148      \\ 
      Number of transactions        & 3,640,961  &  10,025,642  \\
      \hline
    \end{tabular}
  \end{center}
\end{table}

\begin{figure*}
  \hspace*{1cm}
  \includegraphics[width=14cm]{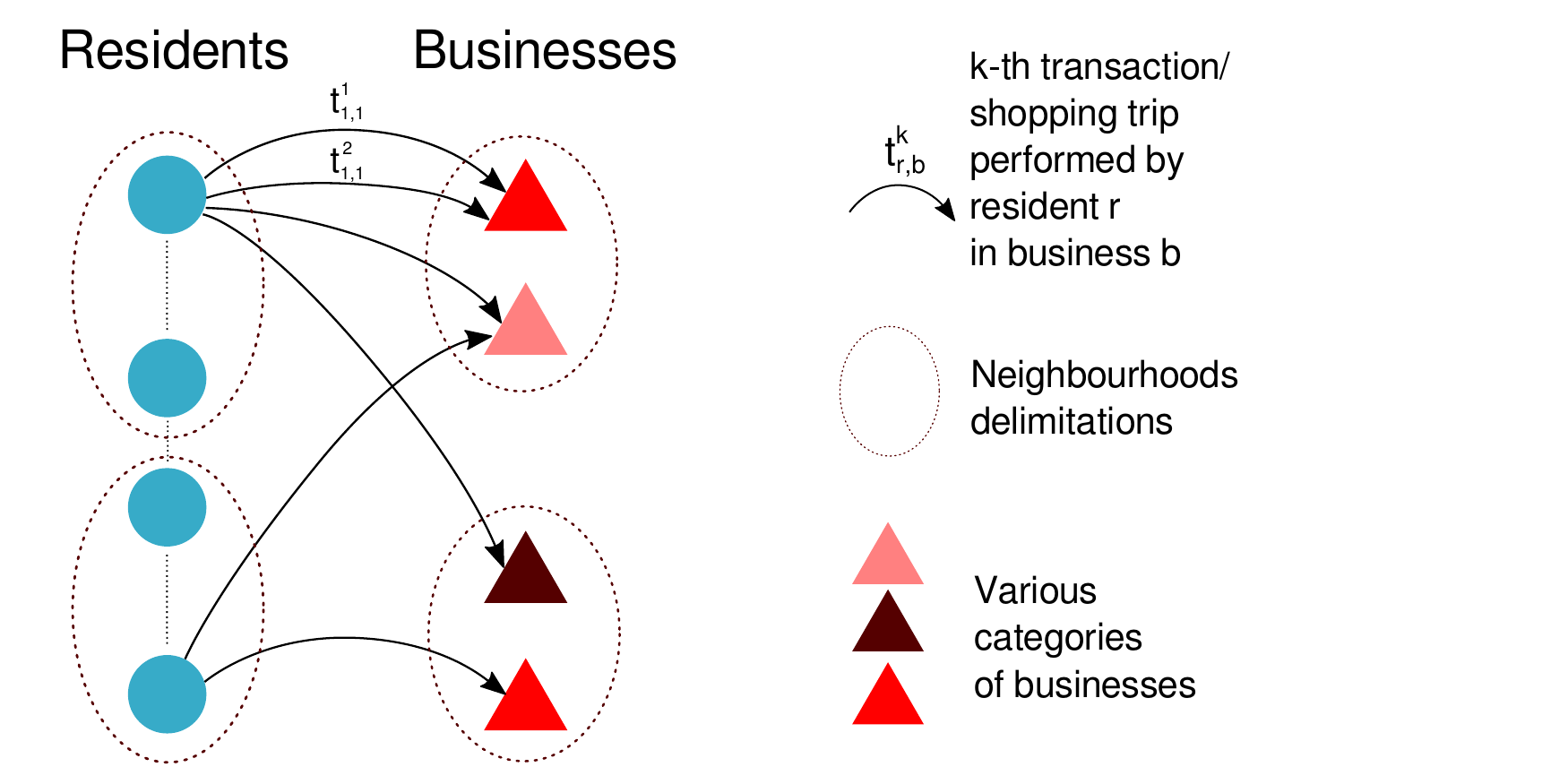}
  \caption{\label{FigS2}\textbf{The bipartite network of transactions.}}
\end{figure*}

The rewiring methods we implemented operate directly at the micro
scale of the individual transactions, and each rewiring operation
$t_{r, b} \rightarrow t_{r, b'}$ consists in selecting a business
$b'\neq b$, such than $b'$ and $b$ are of the same category $c$ but
are located in different neighborhoods. The rewiring occurs only if
$b'$ fulfills a number of additional constraints which are expressed at the macro-scale of the entire city.

The network is rewired iteratively, i.e. transaction per
transaction. A transaction $t_{r,b}$ is picked up randomly (uniform or
weighted sampling, as described in the main text). A neighborhood is
chosen among the set of all neighborhoods that contain some businesses
$b'$ of the same category than $b$. Both the transaction and the
candidate business can be picked up through a uniform or weighted
random sampling. Finally, if the neighborhood change
$j \rightarrow j'$ matches the four constraints, then the transaction/edge is rewired. The
process stops when the rewiring rate falls below $0.001$.

\begin{figure*}[p]
  \centering
  \includegraphics[width=13cm]{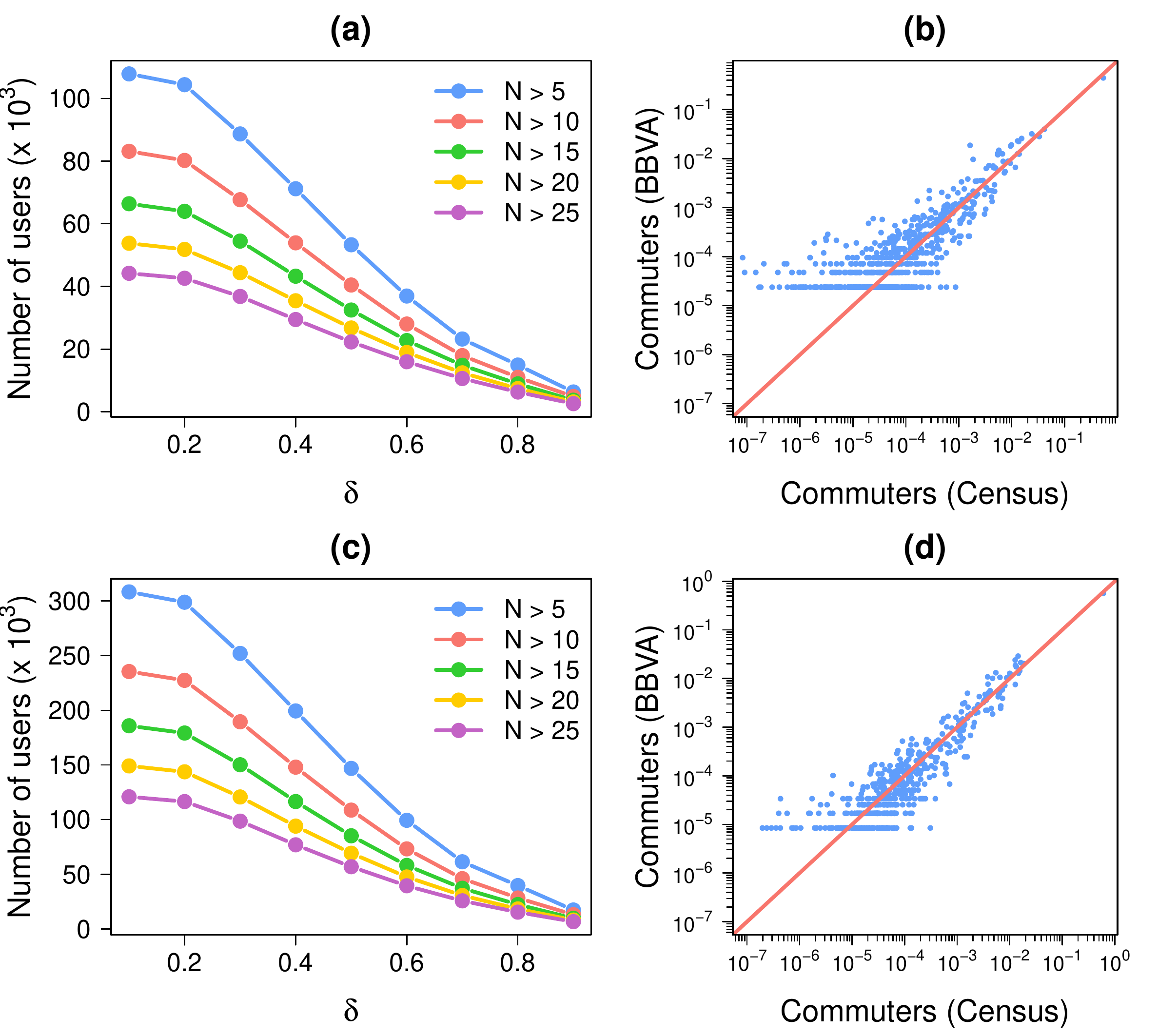}
  \caption{\label{FigS3} \textbf{Identification of the users' main daytime
    activity location in Barcelona ((a)-(b)) and Madrid ((c)-(d)).} (a)
    and (c) Number of users according to $N$ and $\delta$. (b) and (d)
    Comparison between the non-zero flows obtained with the credit
    card dataset ($(N,\delta)=(20,1/3)$) and the census data. The
    values have been aggregated at the municipality scale. The values
    have been normalized by the total number of commuters for both OD
    tables. Blue points are scatter plot for each pair of
    municipalities. The red line represents the x = y line.}
\end{figure*}

\subsection*{Estimation of shopping trips distances and identification
  of the users' main daytime activity location}
\label{sec:distances}

We made the assumption that to each transaction is implicitly
associated a trip originating from either the main activity
neighborhood during working time, or the neighborhood of residence,
depending on the hour of the day and day of the week. The shopping
trip distance is then defined as the Euclidean distance between the
centroids of the origin and destination neighborhoods.

We already know the neighborhood of residence that we can assign as
the place of main activity during night time (i.e between 7pm and 8am)
on week days and Saturday and Sunday. In addition to the neighborhood
of residence, for each individual we can determine the neighborhood in
which he/she was the most frequently located during the typical
working hours of working days, i.e. from 8pm to 7am, from Monday to
Friday.

To do so, for each individual we count the number of unique couples
$(day, hour)$ during which he/she was located in each
neighborhood. For our study we keep only the individuals for which
credit card is a casual mode of payment, and for which we can then
reasonably assume that their card purchases and corresponding shopping
trips are representative of their shopping mobility in
general. Regarding the available statistics for Spain on the share of
credit card payments among all payments, we decided to keep
individuals whose data displayed at least $N=20$ unique couples
$(day, hour)$ during the entire year. For each of these individuals,
we then determine the neighborhood in which they were the most
frequently located during typical working hours. If this neighborhood
accounted for less than one third of the time $\delta=1/3$ in his/her
entire set of locations, then the individual is discarded. As it can
be seen in Figure S\ref{FigS3}a and Figure S\ref{FigS3}c, the couple of
value $(N,\delta)=(20,1/3)$ allow us to keep enough users and discard
the users not showing enough regularity to estimate their main daytime
activity location.

Finally, we can estimate the commuting flows between neighborhoods and
assess the accuracy of the results by comparing these flows with those
obtained from the 2011 Spanish census in Barcelona and
Madrid \footnote{Instituto Nacional de Estad{\'i}stica (National Institute
  for Statistics). Available: http://www.ine.es. Accessed 2016
  April 26.}. The census data is at the municipal level, which
implies that the neighborhoods must be aggregated at the municipality
scale to be able to perform the comparative
analysis. Figure S\ref{FigS3}b and Figure S\ref{FigS3}d show a scattered
plot with the comparison between the flows obtained with the two
matrices. A good agreement between the two ODs is obtained.

The source code of this method is available at \url{https://github.com/maximelenormand/Most-frequented-locations}.

\onecolumngrid
\newpage
\clearpage 

\section*{Supplementary figures}

\begin{figure}[!h]
  \centering
  \includegraphics[width=\linewidth]{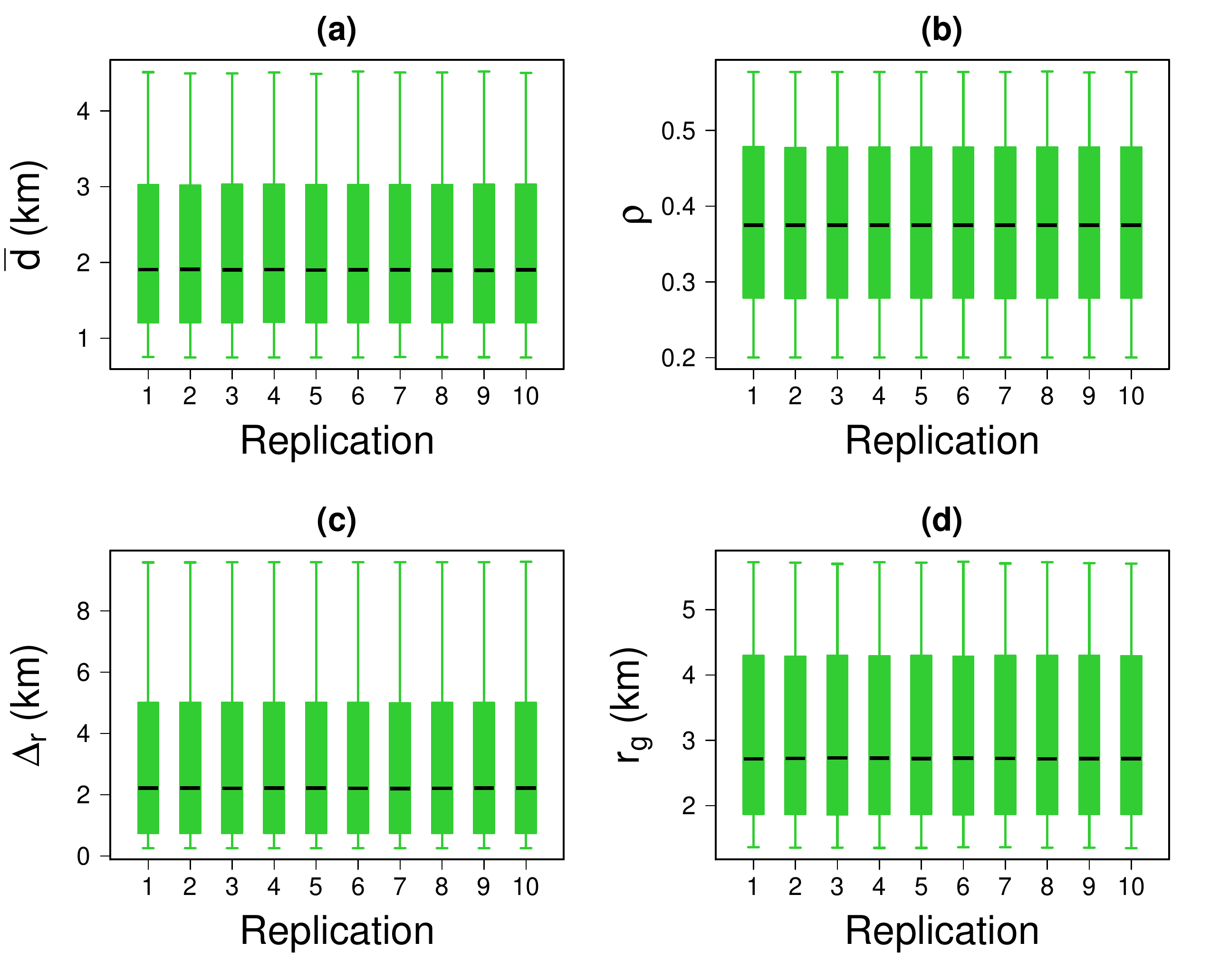}
  \caption{\label{FigS4} \textbf{Individual human mobility indicators'
    distributions obtained with ten replications of the algorithm.}
    (a) Individual average distance traveled $\bar{d}$. (b)
    Exploration rate $\rho$. (c) Jump length distribution
    $\Delta_r$. (d) Radius of gyration $r_g$. The boxplot is composed
    of the first decile, the lower hinge, the median, the upper hinge
    and the 9$^{th}$ decile.}
\end{figure}

\begin{figure}[!h]
  \centering
  \includegraphics[width=\linewidth]{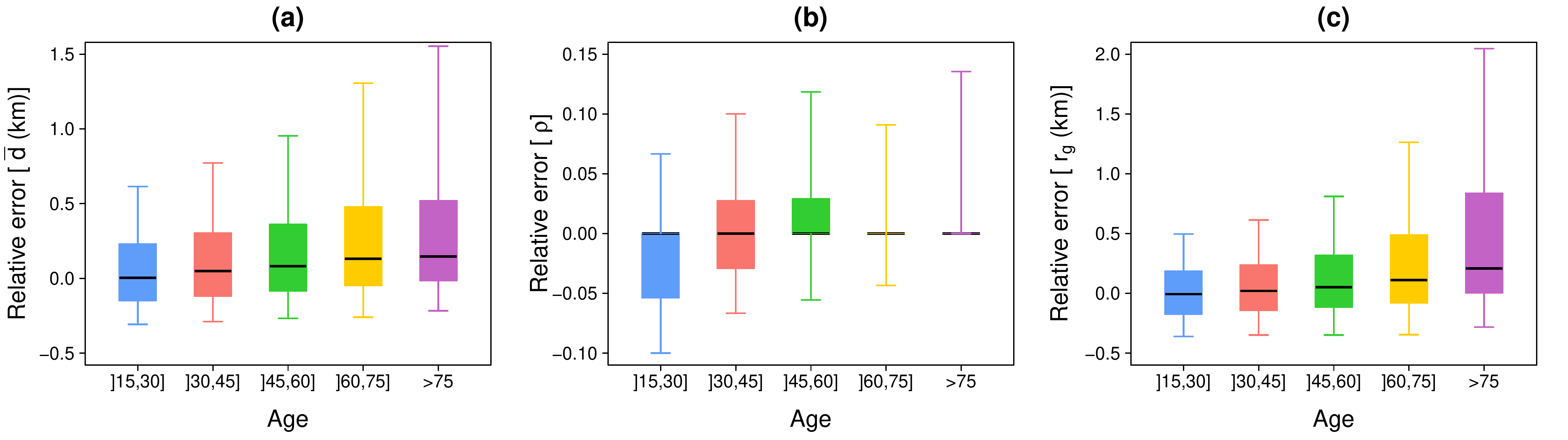}
  \caption{\label{FigS5} \textbf{Relative error between the original users' mobility indicators and the ones obtained after the rewiring according to the age.}
    (a) Individual average distance traveled $\bar{d}$. (b)
    Exploration rate $\rho$. (c) Radius of gyration $r_g$. The relative error is equal to the ratio of the difference between rewiring and original values and the original value. The boxplot is composed of
    the first decile, the lower hinge, the median, the upper hinge and
    the 9$^{th}$ decile.}
\end{figure}

\newpage
\vspace*{4cm}
\begin{figure}[!h]
  \centering
  \includegraphics[width=\linewidth]{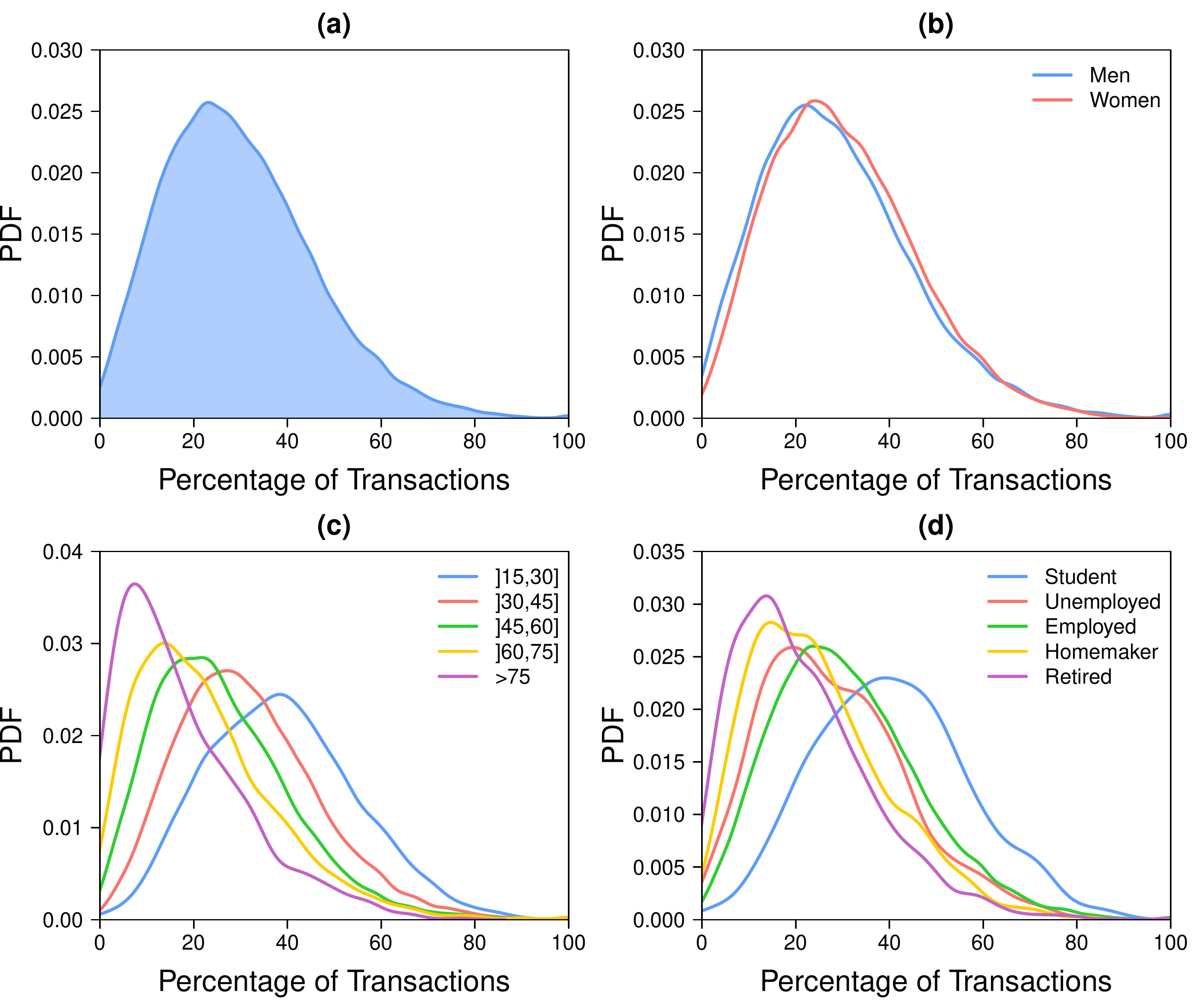}
  \caption{\label{FigS6} \textbf{Probability density functions of the
      individual percentage of rewired transactions.} (a) Total. (b)
    By Gender. (c) By age. (d) By occupation.}
\end{figure}

\newpage
\vspace*{4cm}
\begin{figure}[!h]
  \centering
  \includegraphics[width=\linewidth]{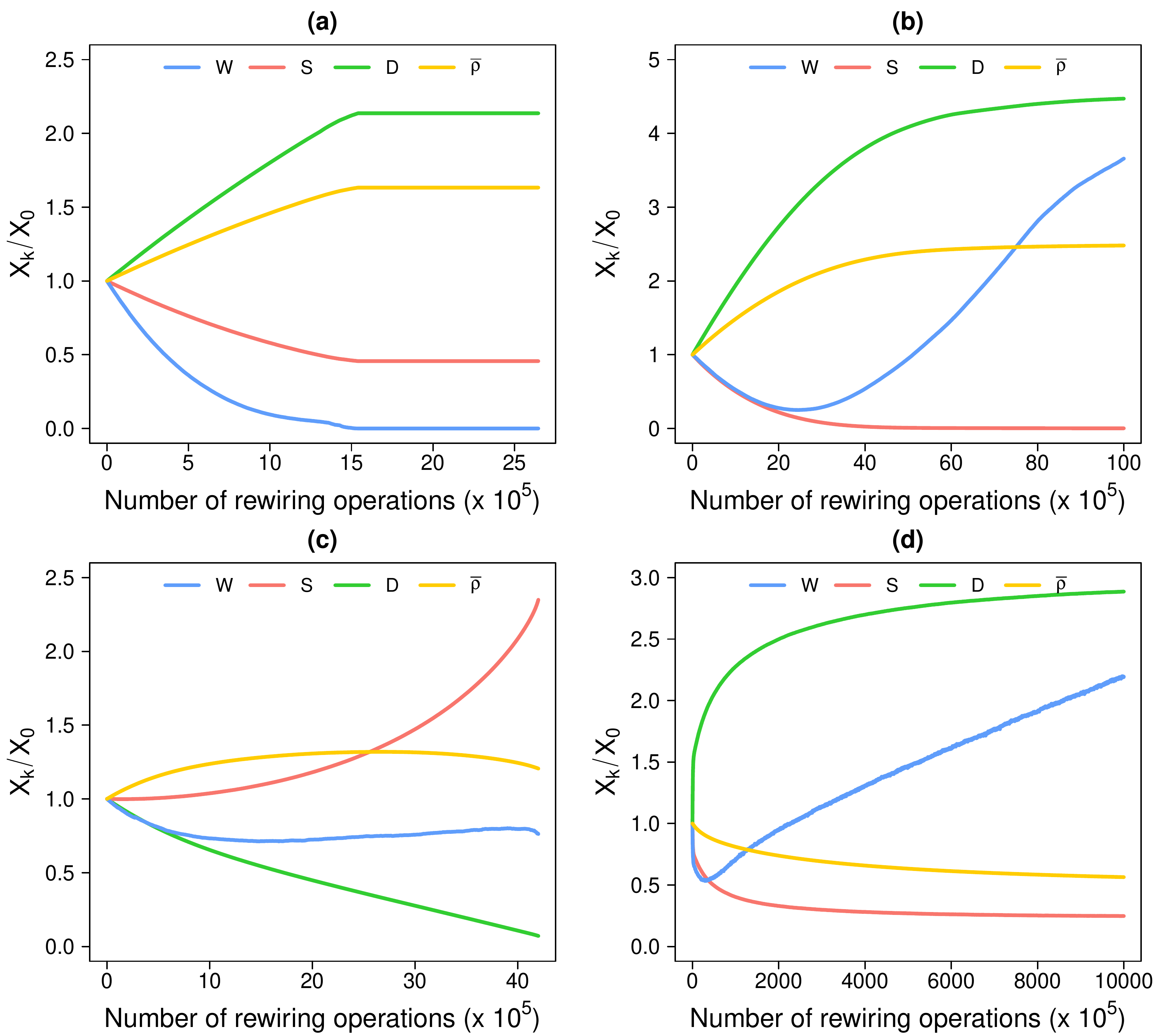}
  \caption{\label{FigS7} \textbf{Evolution of W, S, D and $\bar{\rho}$ as a
    function of the number of rewiring transactions.} (a)
    $(\alpha_W, \alpha_S, \alpha_D,
    \alpha_{\bar{\rho}})=(0,+\infty,+\infty,+\infty)$.
    (b)
    $(\alpha_W, \alpha_S, \alpha_D,
    \alpha_{\bar{\rho}})=(+\infty,0,+\infty,+\infty)$.
    (c)
    $(\alpha_W, \alpha_S, \alpha_D,
    \alpha_{\bar{\rho}})=(+\infty,+\infty,0,+\infty)$.
    (d)
    $(\alpha_W, \alpha_S, \alpha_D,
    \alpha_{\bar{\rho}})=(+\infty,+\infty,+\infty,0)$.}
\end{figure}

\newpage
\vspace*{2cm}
\begin{figure}[!h]
  \centering
  \includegraphics[width=\linewidth]{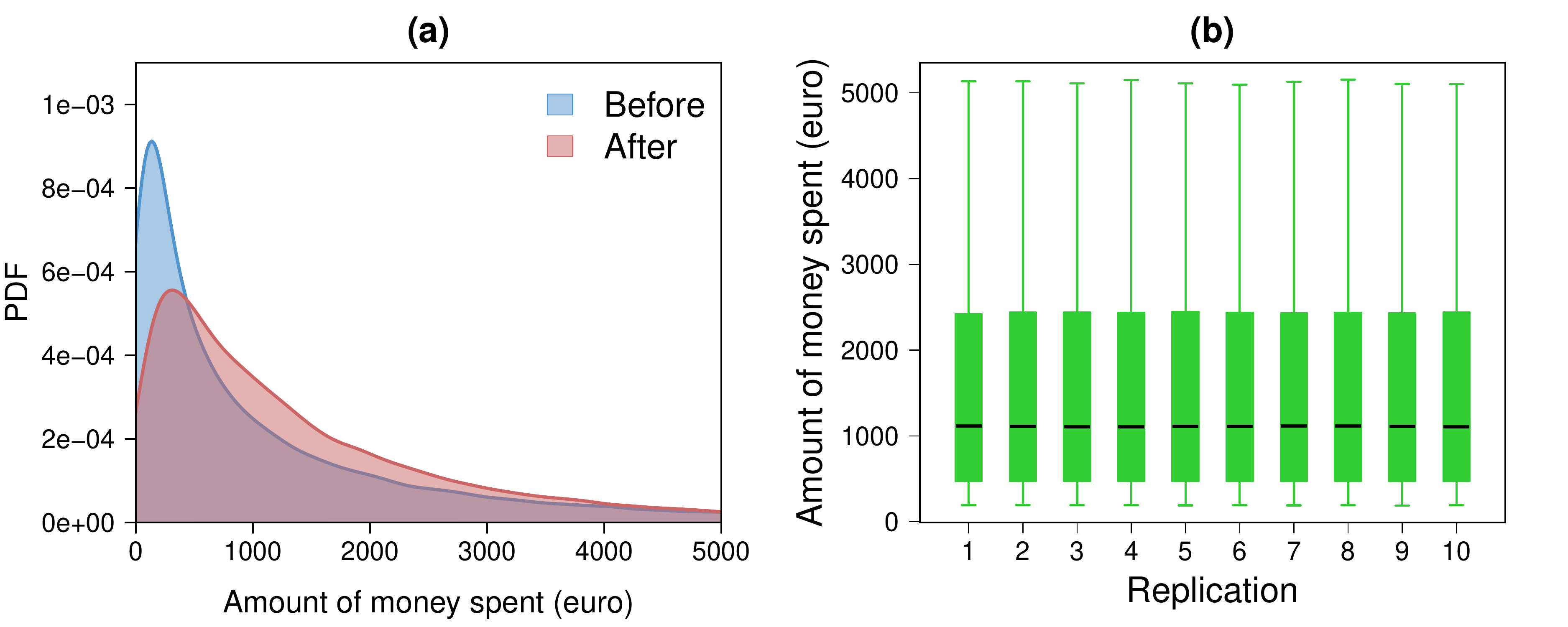}
  \caption{\label{FigS8} \textbf{Probability density functions of the total
    amount of money spent by business in 2011, in Barcelona.} (a)
    Comparison between the original distribution and the one obtained
    after applying the rewiring algorithm. (b) Distributions obtained
    with ten replications of the algorithm. The boxplot is composed of
    the first decile, the lower hinge, the median, the upper hinge and
    the 9$^{th}$ decile. }
\end{figure}

\begin{figure}[!h]
  \centering
  \includegraphics[width=10cm]{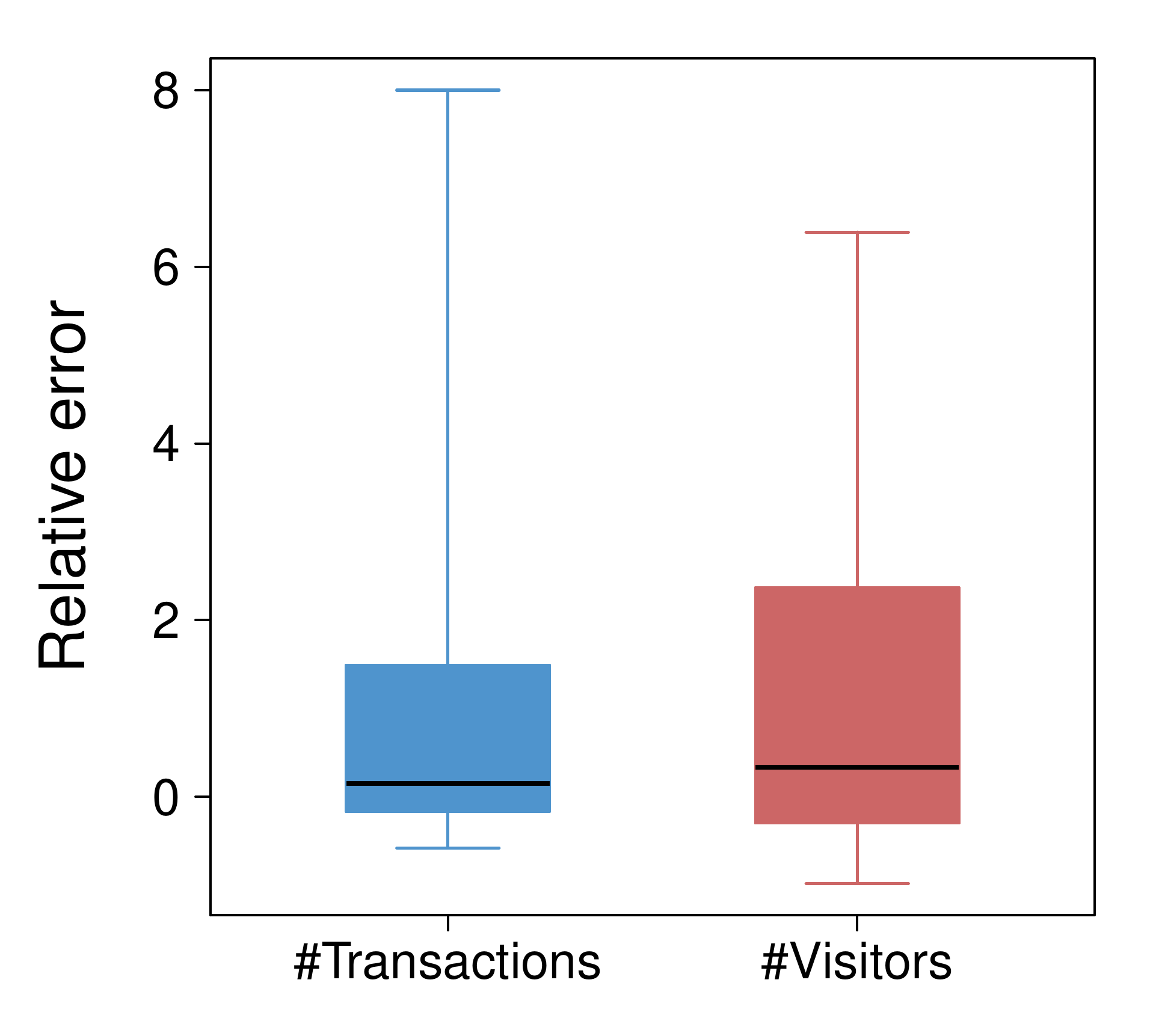}
  \caption{\label{FigS9} \textbf{Relative error between the original number of transactions and visitors and the ones obtained after the rewiring for each business.} The relative error is equal to the ratio of the difference between rewiring and original values and the original value. The boxplot is composed of
    the first decile, the lower hinge, the median, the upper hinge and
    the 9$^{th}$ decile. }
\end{figure}

\newpage
\vspace*{5cm}
\begin{figure}[!h]
  \centering
  \includegraphics[width=12cm]{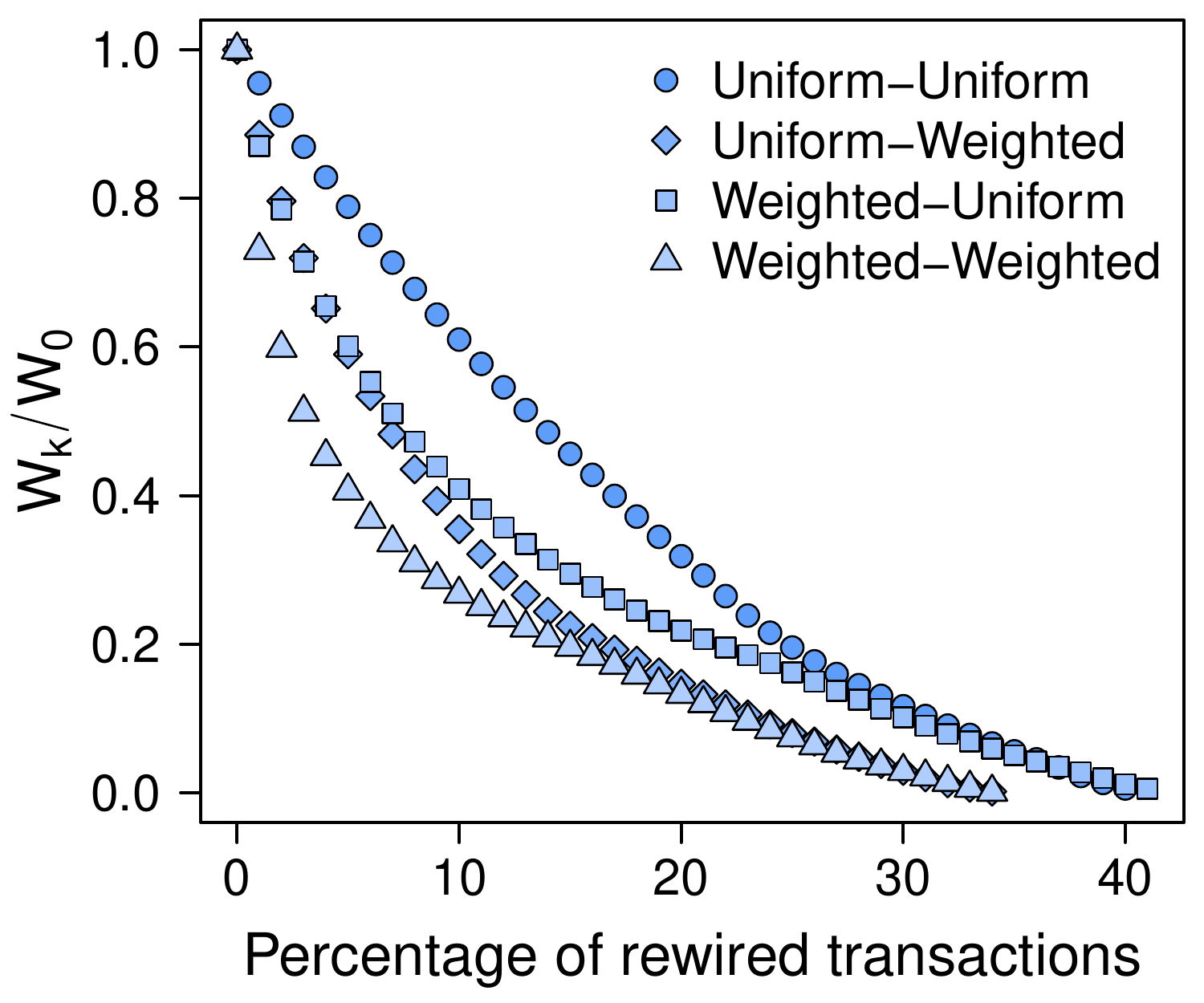}
  \caption{\label{FigS10} \textbf{Decrease of wealth inequality among
    neighborhoods as a function of the fraction of transactions
    rewired, for various heuristics (Madrid).} Four heuristics are
    considered, "Uniform-Uniform", "Uniform-Weighted",
    "Weighted-Uniform" and "Weighted-Weighted".}
\end{figure}

\newpage
\vspace*{5cm}
\begin{figure}[!h]
  \centering
  \includegraphics[width=12cm]{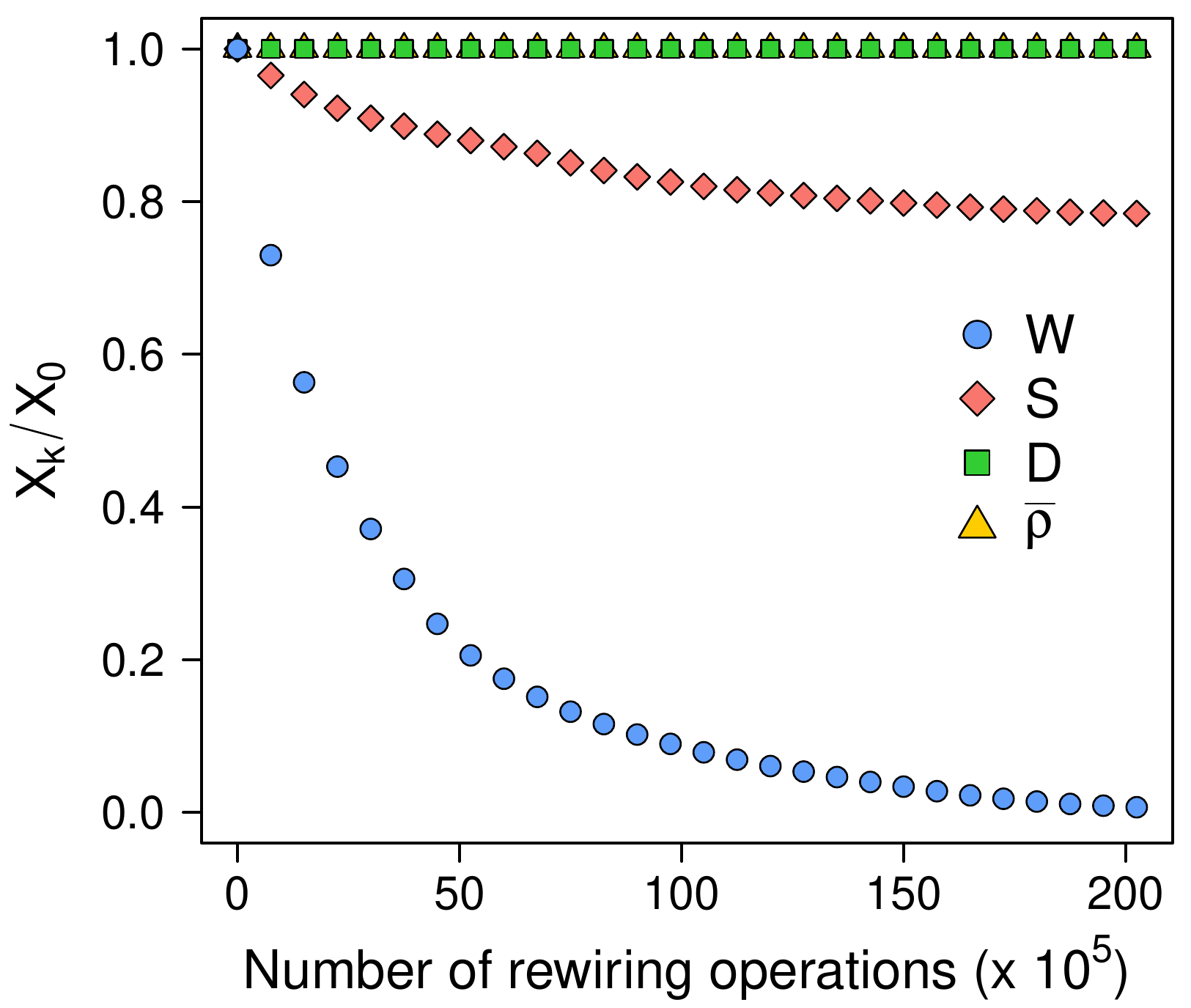}
  \caption{\label{FigS11} \textbf{Decrease of wealth inequality ($W_k/W_0$)
    while preserving the spatial mixing index ($S_k/S_0$), the distance
    traveled ($D_k/D_0$) and the exploration rate
    ($\bar{\rho}_k/\bar{\rho}_0$) as a function of the number of
    rewiring operations (Madrid case).} Values have been averaged over
    hundreds of replications. The bars represent the minimum and the
    maximum values obtained but in most cases they are too close to
    the average to be seen.}
\end{figure}

\newpage
\vspace*{5cm}
\begin{figure}[!h]
  \centering
  \includegraphics[width=\linewidth]{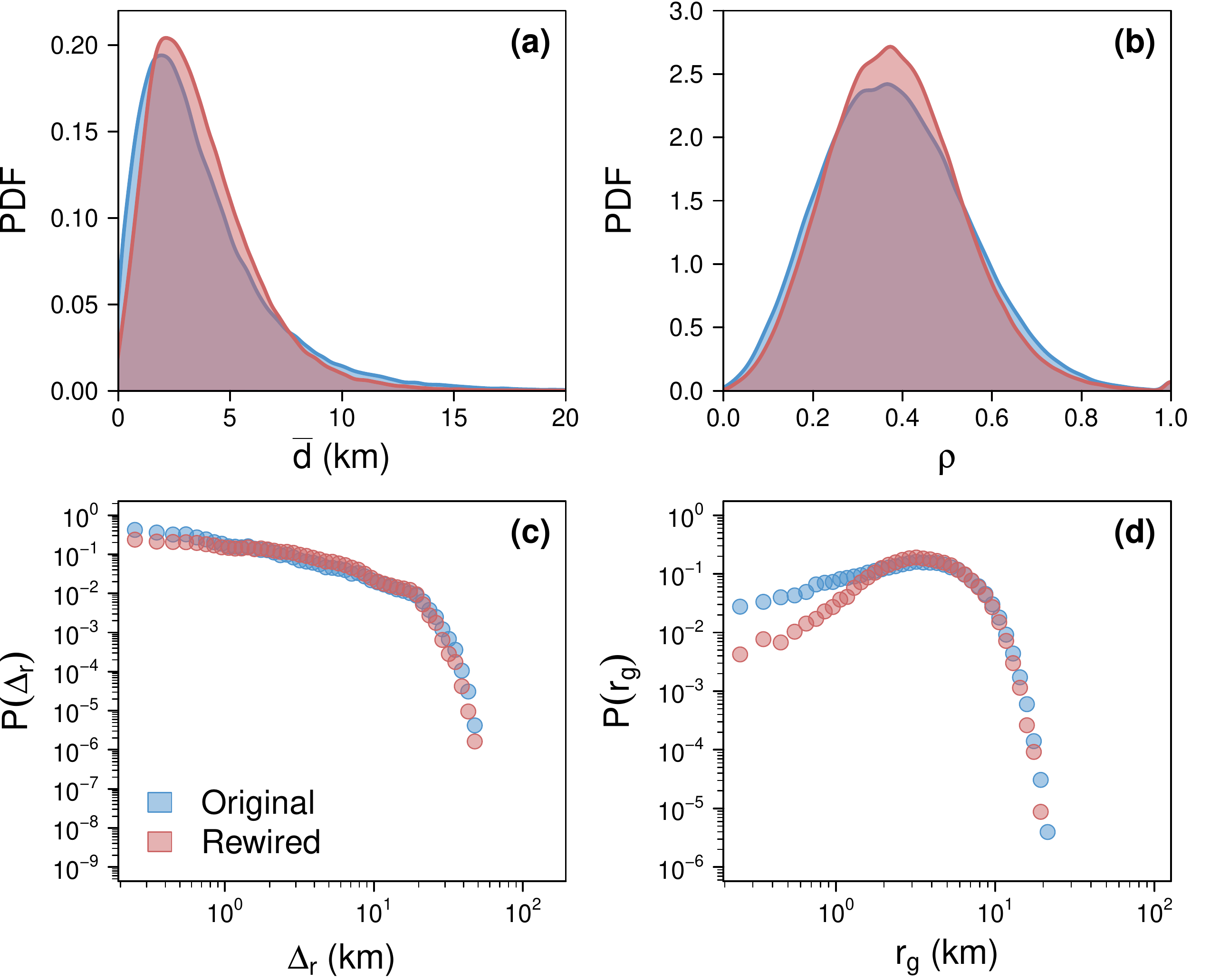}
  \caption{\label{FigS12} \textbf{Observed and simulated distributions of human
    mobility indicators in Madrid.} The distribution of jump lengths
    $\Delta_r$, the radius of gyration $r_g$, the tendency to return
    to already visited places ($\rho$) and the individual average
    distance traveled ($\bar{d}$) are considered. Values measured on
    the empirical data are in blue, while those obtained after
    rewiring are in red. The calculation of $\Delta_r$ and $r_g$ is
    based on the business' exact geographical coordinates.}
\end{figure}

\newpage
\vspace*{5cm}
\begin{figure}[!h]
  \centering
  \includegraphics[width=12cm]{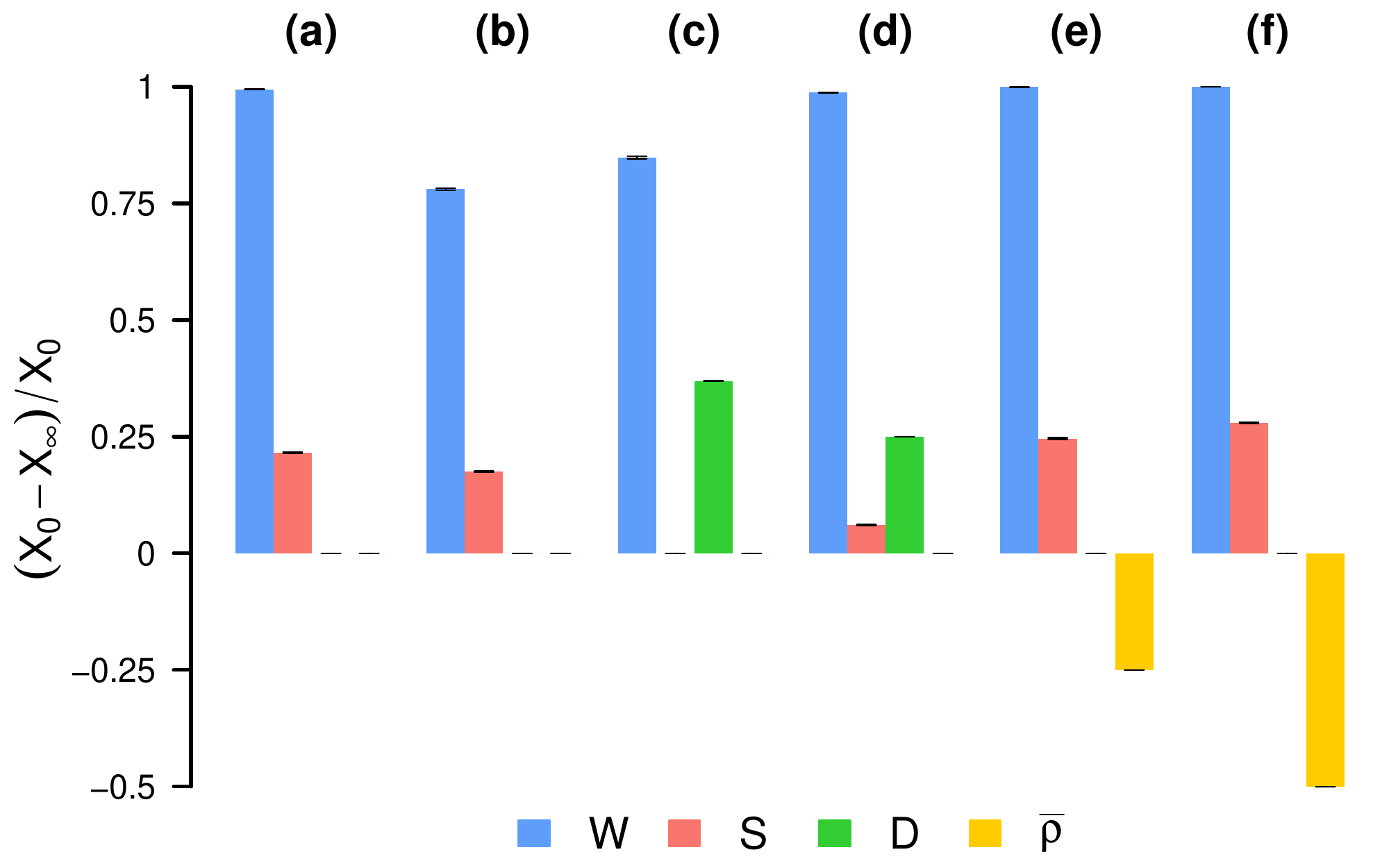}
  \caption{\label{FigS13} \textbf{Multi-criteria improvement of shopping
    mobility in the city of Madrid.} Each group of bars gives the
    relative gains or losses for the four indicators $W$, $S$, $D$ and
    $\bar{\rho}$.}
\end{figure}

\end{document}